\documentclass[%
preprint,superscriptaddress,
nofootinbib,
 amsmath,amssymb,
 aps, prd,
]{revtex4-1}

\usepackage[T1]{fontenc}
\usepackage{graphicx}
\usepackage{dcolumn}
\usepackage{bm}
\usepackage{comment}
\usepackage[dvipsnames,svgnames,x11names]{xcolor}
\usepackage[normalem]{ulem}
\newcommand{\Tr}{\operatorname{Tr}}

\usepackage[colorlinks=true,allcolors=blue]{hyperref}
\begin{document}

\title{Fluctuation--response relations from an emergent
\texorpdfstring{$\mathbb{Z}_2$}{Z2}
symmetry in the rotating stochastic Landau model}

\author{Dhruv Kush}
\email{kush3@illinois.edu}
\affiliation{Illinois Center for Advanced Studies of the Universe and Department of Physics, The Grainger College of Engineering, University of Illinois Urbana-Champaign, Urbana, Illinois 61801, USA}
\author{Nicki Mullins}
\email{nmmulli2@ncsu.edu}
\affiliation{Department of Physics, North Carolina State University, Raleigh, NC 27695, USA}

\author{Mauricio~Hippert}
\email{hippert@cbpf.br}
\affiliation{Centro Brasileiro de Pesquisas F\'isicas, Rua Dr. Xavier Sigaud 150, Rio de Janeiro, RJ, 22290-180, Brazil}

\author{Jorge Noronha} 
\email{jn0508@illinois.edu}
\affiliation{Illinois Center for Advanced Studies of the Universe and Department of Physics, The Grainger College of Engineering, University of Illinois Urbana-Champaign, Urbana, Illinois 61801, USA}

\begin{abstract}
In this work, we investigate the extent to which fluctuation--response relations emerge from coarse-grained stochastic dynamics alone, and which aspects instead depend on additional  information about the system. To address this question, we study the rotating stochastic Landau model, an exactly solvable system describing an overdamped charged Brownian particle in a constant magnetic field, coupled dissipatively to a rotating environment, whose steady state supports circulating probability currents. Using the Martin--Siggia--Rose path integral, we show that there is an emergent $\mathbb{Z}_2$ symmetry transformation that implements the time-reversed dynamics and changes the action by a boundary term. Comparison with the Crooks fluctuation theorem identifies this term with the entropy associated with transitions between steady-state configurations. After coupling the theory to external sources, the same symmetry yields Ward identities relating fluctuations and response. These identities follow entirely from the coarse-grained stochastic theory and do not fix the noise strength. Finally, upon imposing the Einstein relation, we show that they coincide with the high-temperature fluctuation--dissipation relations implied by the rotating Kubo--Martin--Schwinger condition for a microscopic Gibbs ensemble.
\end{abstract}

\date{\today}

\maketitle

\tableofcontents

\section{Introduction and Results}

In many experimental settings, one probes a system only after transient effects subside, making the late-time regime the most directly accessible. Theoretically, this regime is distinguished by the emergence of effective descriptions in which detailed microscopic evolution gives way to a reduced set of relevant degrees of freedom. In this context, effective field theory approaches to late-time dynamics have attracted much attention in recent years; see \cite{Kovtun:2014hpa, Haehl:2014zda, Haehl:2015pja, Sieberer:2015hba, CrossleyGloriosoLiu, Glorioso:2017fpd, GloriosoLiu, JainKovtun, Mullins:2023ott, AndyLucas2023, Mullins:2025vqa}.

When the microscopic system is thermal, the corresponding generating functional obeys Kubo--Martin--Schwinger (KMS) constraints \cite{Glorioso:2017fpd, GloriosoLiu}, which impose fluctuation--dissipation relations \cite{Kubo1966,UHeinzFDT, Chernyak_2006, Chetrite2008Fluctuation} on the effective theory. On the other hand, the Crooks fluctuation theorem \cite{Crooks1998,Crooks1999,JarzynskiWardIdentities_2011} in principle allows analogous identities to be formulated directly at the level of coarse-grained stochastic dynamics, without assuming from the outset that the effective theory descends from a Gibbs state. This was demonstrated in \cite{Mullins:2025vqa}, where a relativistic covariant version of the Crooks fluctuation theorem was shown to give rise to an emergent $\mathbb{Z}_2$ symmetry responsible for imposing fluctuation--dissipation relations for $n$-point correlation functions, similar to what is known to occur in the Schwinger--Keldysh effective action \cite{CrossleyGloriosoLiu, Glorioso:2017fpd, GloriosoLiu}.

It is interesting to note that the two approaches above begin from logically different assumptions. The KMS condition starts from a  thermal state and constrains the effective theory accordingly, whereas a Crooks-based construction starts from a comparison between forward and reverse stochastic trajectories and asks what can be inferred directly from their relative probabilities. This raises a basic question for the late-time behavior of stochastic systems: how much of the familiar relation between fluctuations and response follows from the general structure of the coarse-grained dynamics itself, and how much requires additional knowledge about the system, such as whether it is in thermal equilibrium? This distinction becomes especially subtle when time-reversal-odd backgrounds and steady-state probability currents are present. A useful way to disentangle these ingredients is then to study a model in which the stochastic dynamics, the steady-state probability current, and the response functions can all be computed exactly.

In this paper, we introduce the rotating stochastic Landau model as an exactly solvable benchmark in which the symmetry of the Martin--Siggia--Rose (MSR) action implementing Crooks time reversal, its resulting Ward identities, and their subsequent interpretation in the case of rotating thermal equilibrium can all be displayed explicitly for a steady state with circulating probability currents. We motivate the rotating stochastic Landau model by first describing the conservative theory to which we add dissipation. The planar Landau model describes a charged particle in two dimensions in a harmonic trap and a perpendicular magnetic field \cite{ChernSimonsQM}. Its massless limit is a first-order theory describing a chiral oscillator invariant under global $SO(2)$ rotations and a $\mathbb{Z}_2$ symmetry we dub $\mathsf S\mathsf T$ symmetry (defined by swapping $x$ and $y$ coordinates and applying time reversal). The stochastic theory is then obtained by coupling the conservative theory to a rotating environment. This coupling generates noise and $\mathsf S\mathsf T$-symmetry-breaking dissipative drift terms while preserving $SO(2)$ rotation invariance. Equivalently, this model can be viewed as the overdamped limit of a charged harmonic oscillator coupled to a rotating bath, with a magnetic field applied in the laboratory frame.

We show that the MSR action of this stochastic theory admits a local symmetry transformation that implements time reversal by reversing the direction of the applied magnetic field and the orientation of the rotating medium. The $\mathsf S\mathsf T$-symmetry-breaking drift terms induced by the environment determine the compensating transformation of the MSR response fields under which the stochastic action changes only by a boundary term. Using the Crooks fluctuation theorem \cite{Crooks1998, Crooks1999}, we identify the boundary contribution of this symmetry as the Crooks entropy associated with transitions between steady-state configurations. We then couple the theory to external sources
and obtain the source-only effective action by integrating out the stochastic fields.
Demanding invariance of the effective action under this symmetry transformation leads to MSR Ward identities, derived entirely within the coarse-grained stochastic theory, that relate ``fluctuation'' ($G_S(\omega)$, the symmetric two-point correlator) and response functions ($G_R(\omega)$, the retarded Green function). In deriving these identities, thermal equilibrium or detailed balance is not assumed.

From the Fokker--Planck perspective, the same model describes a Gaussian steady state with a generically non-vanishing probability current. We show that the magnetic field and the rotating environment each contribute independently to the circulating part of the steady-state current and together generate rotational probability flux in the stationary state. Thus, the Ward identities derived from the symmetry that produces the Crooks entropy describe a stochastic theory with rotating probability currents.

We then show how these MSR identities can be derived from a thermal state. A rotating environment naturally leads to a Gibbs state of the form
\begin{equation}
\rho_{\beta,\Omega}
=
\frac{1}{Z}\exp[-\beta(H-\Omega L_z)],
\end{equation}
where $L_z$ is the angular momentum generator and $\Omega$ is the angular velocity of the rotating medium. The corresponding twisted KMS condition \cite{salvio2026thermalgaugetheoryrotating}, in the high-temperature limit, implies a fluctuation--dissipation relation that can be directly matched to our MSR Ward identities by identifying the strength of the noise with an effective temperature of the stochastic theory. Hence, the entropy-generating Crooks symmetry provides response identities, while the KMS matching supplements them with a thermal interpretation in terms of a rotating steady state in thermal equilibrium.

The paper is organized as follows. In Sec.~\ref{sec:conservative-landau}, we review the conservative planar Landau model and its massless first-order limit, and introduce its underlying $\mathsf S\mathsf T$ discrete symmetry. In Sec.~\ref{sec:physical-origin}, we define the stochastic model as describing a rotating Brownian particle with damping, harmonic confinement, and magnetic field, and then take the massless limit. In Sec.~\ref{transition_probabilities}, we construct the corresponding MSR path integral and compute the transition probability. In Sec.~\ref{sec:fokker-planck}, we analyze the stability of the steady state and compute the frequency of probability-current rotation in the steady state from the Fokker--Planck equation. Then, in Sec.~\ref{sec:crooks-general}, the entropy that appears in the Crooks fluctuation theorem is derived from a local symmetry of the MSR action. Subsequently, in Sec.~\ref{sec:src-integrateout}, we couple the theory to sources and derive the associated Ward identities. In Sec.~\ref{app:twisted-kms}, we match the result with the high-temperature limit of the twisted KMS condition for a rotating Gibbs state and interpret the MSR Ward identities in terms of fluctuation--dissipation relations. We present our conclusions and outlook in Sec.~\ref{sec:conclusions}. Details concerning the MSR Jacobian and the Ward identities are provided in the appendices.

\noindent
\emph{Notation:} We use natural units, \(\hbar=c=k_B=1\), and employ the Einstein summation convention for repeated spatial indices \(i,j,\ell=1,2\). We define
\begin{equation}
I_{ij}=\delta_{ij},
\qquad
J_{ij}=\varepsilon_{ij},
\qquad
\varepsilon_{12}=-\varepsilon_{21}=1,
\end{equation}
so that
\begin{equation}
J=
\begin{pmatrix}
0&1\\
-1&0
\end{pmatrix},
\qquad
J^{\top}=-J,
\qquad
J^2=-I.
\end{equation}
Two-dimensional vectors are denoted by bold symbols, for example
\(\bm x=(x_1,x_2)^{\top}\), and \(x_i x_i=\bm x^{\top}\bm x\).
Throughout, a superscript \(\top\) denotes matrix transpose. We denote action functionals by \(\mathcal S\), use \(T=\beta^{-1}\) for temperature and \(\tau=t_f-t_i\) for an elapsed time interval, and reserve \(\omega\) for frequency. Furthermore, we use \(P\) for probability densities and transition probabilities, \(\mathbb P\) for probability functionals, and \(\gamma\) for a stochastic trajectory.

\section{Conservative planar Landau model in the massless limit}
\label{sec:conservative-landau}

We consider the two-dimensional motion of a particle of mass \(m\) and charge \(q\) in  the \(xy\) plane in the presence of an isotropic harmonic potential
\begin{equation}
V(\bm x)=\frac{k}{2}x_i x_i,
\qquad
k>0,
\end{equation}
and subject to a constant and uniform magnetic field \(B\hat{\bm z}\). For the sake of definiteness, we assume $qB>0$. In the symmetric gauge the gauge potential can be written as
\begin{equation}
A_i=-\frac{B}{2}\varepsilon_{ij}x_j,
\end{equation}
and the Lagrangian is
\begin{equation}
L=
\frac{m}{2}\dot x_i\dot x_i
+\frac{qB}{2}\varepsilon_{ij}x_i\dot x_j
-\frac{k}{2}x_i x_i .
\label{eq:Lstart-new}
\end{equation}
The equations of motion can be easily determined to be
\begin{equation}
m\ddot x_i-qB\varepsilon_{ij}\dot x_j+kx_i=0 .
\label{eq:EOM-new}
\end{equation}
Equivalently, to illustrate our notation, they may be written as
\begin{equation}
m\ddot{\bm x}-qB J\dot{\bm x}+k\bm x=0.
\label{eq:EOM-matrix-new}
\end{equation}

The two characteristic frequencies of the system can be found by resolving the motion into circular polarizations. Since \(J=i\sigma_y\), with $\sigma_y$ being the standard corresponding Pauli matrix, its eigenvectors \(\bm e_\varsigma\), with \(\varsigma=\pm1\), satisfy
\begin{equation}
J\bm e_\varsigma=i\varsigma\bm e_\varsigma.
\end{equation}
Substituting
\begin{equation}
\bm x(t)=\bm e_\varsigma e^{-i\omega t}
\end{equation}
into Eq.~\eqref{eq:EOM-matrix-new} then gives
\begin{equation}
m\omega^2+\varsigma qB\,\omega-k=0.
\end{equation}
The two positive normal-mode frequencies are therefore
\begin{equation}
\omega_\pm
=
\bar\omega\pm\frac{qB}{2m},
\qquad
\bar\omega
=
\sqrt{\frac{(qB)^2}{4m^2}+\frac{k}{m}},
\label{eq:fock-darwin-frequencies}
\end{equation}
and they obey the following relation
\begin{equation}
\omega_+\omega_-=\frac{k}{m}.
\label{eq:frequency-product}
\end{equation}
These normal modes describe circular motions of opposite chirality.

We now take the low-energy limit in which \(m\to0\), while \(qB\neq0\) and \(k\) are kept fixed. One can see that
\begin{equation}
\omega_+
=
\frac{qB}{m}
+\frac{k}{qB}
+O(m),
\qquad
\omega_-
=
\frac{k}{qB}
+O(m),
\label{eq:frequencies-small-m-positive}
\end{equation}
and, thus, in this limit the high-frequency cyclotron mode is projected out and only one chiral mode survives with frequency $\omega_{-}$.

The physics of this system can, of course, also be understood directly from a Lagrangian. In the limit \(m\to0\), the inertial term disappears and Eq.~\eqref{eq:Lstart-new} reduces to the first-order Lagrangian describing a chiral oscillator studied in \cite{ChernSimonsQM}
\begin{equation}
L_0=
\frac{qB}{2}\varepsilon_{ij}x_i\dot x_j
-\frac{k}{2}x_i x_i 
\label{eq:Lmassless-new}
\end{equation}
with equation of motion
\begin{equation}
-qB J\dot{\bm x}+k\bm x=0.
\label{eq:planar-landau-eom}
\end{equation}
Multiplying this equation by \(J\) and using \(J^2=-I\), one obtains
\begin{equation}
\dot{\bm x}
=
-\frac{k}{qB}J\bm x.
\label{eq:massless-rotation}
\end{equation}
To understand the motion of the particle in this case, we note that
\begin{equation}
\frac{d}{dt}\left(\bm x^{\top}\bm x\right)
=
2\bm x^{\top}\dot{\bm x}
=
-\frac{2k}{qB}\bm x^{\top} J\bm x
=
0.
\end{equation}
Thus, the distance from the origin is conserved, i.e., the surviving degree of freedom therefore moves on a circle with angular frequency \(k/qB\).

It is also instructive to express the first-order character of this theory in Hamiltonian language, see \cite{ChernSimonsQM}. From Eq.~\eqref{eq:Lmassless-new}, the symplectic two-form on the reduced phase space is
\begin{equation}
\varpi=qB\,dx_1\wedge dx_2
\end{equation}
and its inverse determines the reduced Poisson brackets,
\begin{equation}
\{x_i,x_j\}
=
-\frac{1}{qB}\varepsilon_{ij}
\label{eq:reduced-poisson-bracket}
\end{equation}
and the corresponding reduced Hamiltonian
\begin{equation}
H_0=\frac{k}{2}x_i x_i .
\label{eq:reduced-hamiltonian}
\end{equation}
Hamilton's equation then gives
\begin{equation}
\dot x_i
=
\{x_i,H_0\}
=
-\frac{k}{qB}\varepsilon_{ij}x_j,
\end{equation}
in agreement with Eq.~\eqref{eq:massless-rotation}.

It is useful to discuss here the discrete symmetries of this system. Let \(\mathsf S\) denote the spatial reflection that exchanges the two Cartesian coordinates,
\begin{equation}
\mathsf S=\sigma_x
=
\begin{pmatrix}
0&1\\
1&0
\end{pmatrix},
\qquad
\mathsf S^{\top}=\mathsf S=\mathsf S^{-1}.
\end{equation}
This is an orientation-reversing reflection satisfying
\begin{equation}
\mathsf S^{\top}J\mathsf S=-J .
\end{equation}
We denote by \(\mathsf T\) the usual time-reversal operation on the particle's trajectory,
\begin{equation}
\mathsf T:\qquad
\bm x(t)\to \bm x(-t),
\qquad
\dot{\bm x}(t)\to -\dot{\bm x}(-t).
\end{equation}
At fixed nonzero \(B\), neither \(\mathsf T\) nor \(\mathsf S\) is separately a symmetry, because each reverses the sign of the magnetic term
\(\bm x^{\top} J\dot{\bm x}\). Their product, however, leaves this term invariant. Thus, the combined transformation acts as
\begin{equation}
\mathsf S\mathsf T:\qquad
\bm x(t)\to \mathsf S\bm x(-t),
\qquad
\dot{\bm x}(t)\to -\mathsf S\dot{\bm x}(-t),
\label{eq:ST-transformation}
\end{equation}
under which
\begin{equation}
\dot{\bm x}^{\top}\dot{\bm x},
\qquad
\bm x^{\top} J\dot{\bm x},
\qquad
\bm x^{\top}\bm x
\end{equation}
are all invariant. Therefore, both the massive conservative Landau model and its first-order \(m\to0\) reduction are invariant under \(SO(2)\) rotations and under the discrete \(\mathsf S\mathsf T\) symmetry.
We denote this combined operation by \(\Theta\equiv\mathsf S\mathsf T\).

Finally, we note that the magnetic field changes sign under both time reversal $\mathsf T$ and the orientation-reversing spatial reflection $\mathsf S$,
so that the combined transformation leaves the magnetic field unchanged. We will see below that the stochastic theory retains the first-order rotational structure of the reduced Landau model, while the rotating bath introduces additional dissipative and stochastic terms whose behavior under these transformations must be analyzed separately.

\section{Defining the rotating stochastic Landau model}
\label{sec:physical-origin}

We now define the rotating stochastic Landau model as describing the overdamped dynamics of a charged Brownian particle coupled to a rotating bath \cite{BrownianBfield, Sahoo, Pradhan}. The stochastic differential equation derived below combines the first-order structure inherited from the closed \(\mathsf S\mathsf T\)-symmetric planar Landau model with two additional bath-induced, \(\mathsf S\mathsf T\)-breaking terms generated by dissipation. 

We start with a massive charged Brownian particle in a two-dimensional harmonic potential, with stiffness \(K>0\), coupled to a bath that is rotating in the laboratory frame with frequency $\Omega$, in the absence of a magnetic field. Let the laboratory-frame coordinates of the particle be \(\bm x(t)\), and let \(\bm y(t)\) denote its coordinates in the co-rotating frame of the bath. The stochastic differential equation describing this Brownian particle in the co-rotating frame is
\begin{equation}
m\ddot{\bm y}
-
m\Omega \,J \dot{\bm y}
-
m\Omega^2 {\bm y}
+
\Gamma \dot{\bm y}
+
K\bm y
=
\bm\xi(t)
\label{eq:massive-stochastic-landau-co-rotating}
\end{equation}
where $\Gamma \dot{\bm y}$ denotes the ordinary viscous drag (with damping coefficient $\Gamma>0$), the second and third terms are due to inertial effects, and we represent the random force exerted by the bath by an isotropic Gaussian white noise \(\bm\xi(t)\) satisfying
\begin{equation}
\left\langle \xi_i(t)\right\rangle=0,
\qquad
\left\langle \xi_i(t)\xi_j(t')\right\rangle
=
2D_{\rm B}\delta_{ij}\delta(t-t'),
\qquad
D_{\rm B}>0.
\label{eq:predivision-noise}
\end{equation}
As in the previous section, we will consider here the limit where $m\to 0$, which gives
\begin{equation}
\Gamma \dot{\bm y}
+
K\bm y
=
\bm\xi(t).
\label{eq:massive-stochastic-landau-co-rotating-m0}
\end{equation}
We now write this equation in the laboratory frame using that 
\begin{equation}
\bm x(t)=R(t)\bm y(t),
\qquad
R(t)=e^{-\Omega Jt}
\label{eq:bath-rotation}
\end{equation}
so
\begin{equation}
R\dot{\bm y}
=
\dot{\bm x}+\Omega J\bm x.
\label{eq:relative-velocity}
\end{equation}
Thus, the dissipative force \(-\Gamma\dot{\bm y}\) in the co-rotating frame of the bath becomes
\begin{equation}
-\Gamma\left(\dot{\bm x}+\Omega J\bm x\right)
\label{eq:rotating-drag}
\end{equation}
in the inertial laboratory frame. To fully specify the system we are interested in, we now apply a magnetic field \(B_{\rm phys}\hat{\bm z}\) in the laboratory frame so that the stochastic differential equation describing our Brownian particle in the $m\to 0$ is therefore
\begin{equation}
\left(\Gamma I-qB_{\rm phys}J\right)\dot{\bm x}
+
(K I 
+
\Gamma\Omega J)\bm x
=
\bm\xi(t).
\label{eq:massless-stochastic-landau}
\end{equation}
Here, the term \(K\bm x\) arises from the harmonic potential, \(\Gamma\dot{\bm x}\) is the ordinary viscous drag, \(-qB_{\rm phys}J\dot{\bm x}\) is the corresponding Lorentz-force contribution, and \(\Gamma\Omega J\bm x\) appears because dissipation acts on the velocity relative to the rotating bath (the other inertial terms proportional to $m$ vanish in the limit considered here). Finally, we note that the form of the noise is preserved under the time-dependent orthogonal transformation between the co-rotating and laboratory coordinates (we use the same variable $\bm\xi$ for the sake of simplicity).
The SDE in \eqref{eq:massless-stochastic-landau}, describing an overdamped charged Brownian particle coupled to a rotating bath, is what we refer to as the rotating stochastic Landau model in this work.

Comparing with Eq.~\eqref{eq:planar-landau-eom}, we remark that this equation of motion contains the \(\mathsf S\mathsf T\)-symmetric terms proportional to \(J\dot{\bm x}\) and \(I\bm x\) descending from the conservative formulation. Crucially, it also contains \(\mathsf S\mathsf T\)-symmetry-breaking dissipative terms proportional to \(I\dot{\bm x}\) and \(J\bm x\) arising from the rotating bath. Indeed, under the fixed-background \(\mathsf S\mathsf T\) transformation defined in Eq.~\eqref{eq:ST-transformation}, the terms \(J\dot{\bm x}\) and \(I\bm x\) transform covariantly, whereas \(I\dot{\bm x}\) and \(J\bm x\) acquire an additional minus sign.

\section{Transition probabilities in the rotating stochastic Landau model}
\label{transition_probabilities}

\subsection{Derivation of the MSR path integral}

For the sake of completeness, we review the derivation of the MSR path integral \cite{MSR}. Consider a real random variable \(x(t)\) undergoing stochastic dynamics. Given an initial condition \(x(0)=x_i\), we are interested in the transition probability that the system attains the value \(x(t_f)=x_f\) at final time \(t_f\). This can be written as
\begin{equation}
P\!\left(x_f,t_f\mid x_i,0\right)
=
\int\mathcal D\xi\,\mathbb P_\xi[\xi]\,
\delta\!\left(x_\xi(t_f)-x_f\right),
\label{eq:transition-prob-noise}
\end{equation}
where \(x_\xi(t)\) is the solution of the corresponding Langevin equation for the noise realization \(\xi(t)\) and probability functional $\mathbb P_\xi[\xi]$, with \(x_\xi(0)=x_i\). The delta function ensures that only those noise realizations that evolve \(x_i\) to \(x_f\) contribute to the transition probability.

Let the Langevin equation be written as
\begin{equation}
F[x](t)=\xi(t).
\label{eq:generic-langevin-F}
\end{equation}
For a causal discretization and a fixed initial value, the map between the noise history and the stochastic trajectory is implemented by the functional identity
\begin{equation}
1
=
\int_{x(0)=x_i}\mathcal Dx\,
\delta\!\left[F[x]-\xi\right]\,
\Delta[x],
\label{deltafn}
\end{equation}
where
\begin{equation}
\Delta[x]
=
\det\!\left[
\frac{\delta F[x](t)}
{\delta x(t')}
\right]
\label{eq:generic-msr-jacobian}
\end{equation}
is the corresponding functional Jacobian. The final endpoint is imposed separately by the delta function in Eq.~\eqref{eq:transition-prob-noise}. For an \(n\)-component variable \(\bm x(t)\), the corresponding identity is
\begin{equation}
1
=
\int_{\bm x(0)=\bm x_i}\mathcal D\bm x\,
\prod_{a=1}^{n}
\delta\!\left[F_a[\bm x]-\xi_a\right]\,
\Delta[\bm x].
\end{equation}
Using the Fourier representation of the delta functional, one obtains
\begin{equation}
\delta\!\left[F[x]-\xi\right]
=
\int\mathcal D\hat x\,
\exp\!\left[
i\int_0^{t_f}dt\,
\hat x(t)\bigl(F[x](t)-\xi(t)\bigr)
\right],
\label{eq:delta-functional-fourier}
\end{equation}
where the response field \(\hat x\) is initially integrated along a real Fourier contour. Inserting Eqs.~\eqref{deltafn} and \eqref{eq:delta-functional-fourier} into Eq.~\eqref{eq:transition-prob-noise} gives
\begin{align}
P(x_f,t_f\mid x_i,0)
&=
\int\mathcal D\xi\,\mathbb P_\xi[\xi]
\int_{x(0)=x_i}^{x(t_f)=x_f}\mathcal Dx
\int\mathcal D\hat x\,
\Delta[x]
\nonumber\\
&\qquad\times
\exp\!\left[
i\int_0^{t_f}dt\,
\hat x(t)\bigl(F[x](t)-\xi(t)\bigr)
\right].
\label{eq:path-int-P}
\end{align}

In this work, we take the noise to be Gaussian with zero mean and symmetric positive covariance kernel \(Q\),
\begin{equation}
\mathbb P_\xi[\xi]
=
\mathcal N_Q
\exp\!\left[
-\frac{1}{2}
\int dt\,dt'\,
\xi(t)Q^{-1}(t,t')\xi(t')
\right].
\label{eq:generic-gaussian-noise}
\end{equation}
The normalized Gaussian average gives
\begin{align}
&\int\mathcal D\xi\,\mathbb P_\xi[\xi]\,
\exp\!\left[
-i\int dt\,\hat x(t)\xi(t)
\right]
\nonumber\\
&\qquad=
\exp\!\left[
-\frac{1}{2}
\int dt\,dt'\,
\hat x(t)Q(t,t')\hat x(t')
\right]
\label{eq:generic-noise-average}
\end{align}
and, consequently,
\begin{align}
P(x_f,t_f\mid x_i,0)
&=
\mathcal N
\int_{x(0)=x_i}^{x(t_f)=x_f}
\mathcal Dx\,\mathcal D\hat x\,
\Delta[x]
\nonumber\\
&\qquad\times
\exp\!\left[
i\int dt\,\hat x(t)F[x](t)
-\frac{1}{2}
\int dt\,dt'\,
\hat x(t)Q(t,t')\hat x(t')
\right].
\label{eq:GN-int}
\end{align}
Integrating over the response field yields the Onsager--Machlup representation
\begin{align}
P(x_f,t_f\mid x_i,0)
&=
\mathcal N
\int_{x(0)=x_i}^{x(t_f)=x_f}
\mathcal Dx\,
\Delta[x]\,
e^{\mathcal S_{\rm OM}[x]},
\label{eq:GN-result}
\\
\mathcal S_{\rm OM}[x]
&=
-\frac{1}{2}
\int dt\,dt'\,
F[x](t)Q^{-1}(t,t')F[x](t').
\label{eq:generic-OM-action}
\end{align}

Thus, the MSR path integral depends on the deterministic Langevin operator, the normalized noise measure, and the Jacobian in Eq.~\eqref{eq:generic-msr-jacobian}. Since the Langevin equation considered below is linear, the Onsager--Machlup action is quadratic and the path integral is Gaussian. Its endpoint dependence is therefore determined exactly by the classical path, while the Gaussian fluctuation determinant fixes the normalization. Computing the functional representation also requires a consistent discretization of the stochastic equation and its Jacobian. We use the midpoint, or Stratonovich, prescription when evaluating the continuum Jacobian, which is done in Appendix~\ref{app:msr-jacobian}. 

\subsection{Computing the MSR path integral}

We now compute the MSR path integral associated with the transition probabilities of the rotating stochastic Landau model. The overdamped equation derived in Sec.~\ref{sec:physical-origin} is
\begin{equation}
\Gamma\dot{\bm x}
-qB_{\rm phys}J\dot{\bm x}
+K\bm x
+\Gamma\Omega J\bm x
=
\bm\xi(t).
\label{eq:overdamped-rotating-B-repeat}
\end{equation}
It is convenient to divide the equation by \(\Gamma\), which gives
\begin{equation}
\left(
I-\frac{qB_{\rm phys}}{\Gamma}J
\right)\dot{\bm x}
+
\Omega J\bm x
+
\frac{K}{\Gamma}\bm x
=
\bm\eta(t),
\qquad
\bm\eta(t)=\frac{\bm\xi(t)}{\Gamma}.
\label{eq:normalized-brownian-sde}
\end{equation}
We note that, since
\begin{equation}
\left\langle
\xi_i(t)\xi_j(t')
\right\rangle
=
2D_{\rm B}\delta_{ij}\delta(t-t'),
\end{equation}
then
\begin{equation}
\left\langle
\eta_i(t)\eta_j(t')
\right\rangle
=
2D\delta_{ij}\delta(t-t'),
\qquad
D=\frac{D_{\rm B}}{\Gamma^2}.
\label{eq:rescaled-noise}
\end{equation}
In what follows, \(D\) denotes this rescaled noise strength.

The transition probability can be represented as
\begin{align}
P(\bm x_f,t_f\mid\bm x_i,t_i)
&=
\Delta
\int_{\bm x(t_i)=\bm x_i}^{\bm x(t_f)=\bm x_f}
\mathcal D\bm x\,
\nonumber\\
&\qquad\times
\left\langle
\delta\!\left[
\left(
I-\frac{qB_{\rm phys}}{\Gamma}J
\right)\dot{\bm x}
+\Omega J\bm x
+\frac{K}{\Gamma}\bm x
-\bm\eta
\right]
\right\rangle_{\eta}.
\label{eq:transition-prob-delta-brownian}
\end{align}
Here, \(\langle\cdots\rangle_\eta\) denotes the normalized Gaussian average over the zero-mean noise \(\bm\eta\) with covariance given in Eq.~\eqref{eq:rescaled-noise}.
Introducing the MSR response field \(\bar{\bm x}\) gives
\begin{equation}
P(\bm x_f,t_f\mid\bm x_i,t_i)
=
\mathcal N\Delta
\int_{\bm x(t_i)=\bm x_i}^{\bm x(t_f)=\bm x_f}
\mathcal D\bm x\,\mathcal D\bar{\bm x}\,
e^{\mathcal S_{\rm MSR}[\bm x,\bar{\bm x}]},
\label{eq:msr-path-integral-brownian}
\end{equation}
where
\begin{equation}
\mathcal S_{\rm MSR}
=
\int_{t_i}^{t_f}dt\,
\left\{
i\bar{\bm x}^{\top}
\left[
\left(
I-\frac{qB_{\rm phys}}{\Gamma}J
\right)\dot{\bm x}
+\Omega J\bm x
+\frac{K}{\Gamma}\bm x
\right]
-
D\,\bar{\bm x}^{\top}\bar{\bm x}
\right\}.
\label{eq:msr-action-brownian}
\end{equation}
Here, \(\Delta\) is the Jacobian associated with the linear Langevin operator. For the additive linear process considered here, this Jacobian is independent of the trajectory and is included in the overall normalization, see Appendix~\ref{app:msr-jacobian}.

The bath-induced terms are visible directly in Eq.~\eqref{eq:msr-action-brownian}. The conservative massless Landau structure is carried by the terms proportional to
\(i\bar{\bm x}^{\top}J\dot{\bm x}\) and
\(i\bar{\bm x}^{\top}\bm x\), while the dissipative coupling to the rotating bath produces the additional structures
\(i\bar{\bm x}^{\top}\dot{\bm x}\) and
\(i\bar{\bm x}^{\top}J\bm x\).
These two terms break the fixed-background \(\mathsf S\mathsf T\) symmetry of the conservative parent theory. Indeed, extending the operation \(\Theta\) to the MSR response field gives
\begin{equation}
\Theta:\qquad
\bm x(t)\to \mathsf S\bm x(-t),
\qquad
\bar{\bm x}(t)\to \mathsf S\bar{\bm x}(-t),
\qquad
\mathsf S^{\top}J\mathsf S=-J,
\label{eq:ST-MSR-fields}
\end{equation}
and one finds
\begin{equation}
\bar{\bm x}^{\top}\dot{\bm x}
\to
-\bar{\bm x}^{\top}\dot{\bm x},
\qquad
\bar{\bm x}^{\top}J\bm x
\to
-\bar{\bm x}^{\top}J\bm x.
\label{eq:ST-breaking-bath-terms}
\end{equation}
Thus, these terms break the \(\mathsf S\mathsf T\) symmetry of the conservative theory while preserving global \(SO(2)\) rotational invariance.

Integrating out the response field gives
\begin{equation}
\mathcal S_{\rm OM}[\bm x]
=
-\frac{1}{4D}
\int_{t_i}^{t_f}dt\,
(\mathcal L\bm x)^{\top}(\mathcal L\bm x),
\label{eq:Seff-action-start-brownian}
\end{equation}
where
\begin{equation}
\mathcal L
=
\left(
I-\frac{qB_{\rm phys}}{\Gamma}J
\right)\partial_t
+\Omega J
+\frac{K}{\Gamma}I.
\label{eq:def-L-brownian}
\end{equation}
It is useful to factorize this operator as
\begin{equation}
\mathcal L
=
\left(
I-\frac{qB_{\rm phys}}{\Gamma}J
\right)
\left(
\partial_t+sI+rJ
\right),
\label{eq:L-factorized-brownian}
\end{equation}
where
\begin{equation}
s
=
\frac{\Gamma\left(K-qB_{\rm phys}\Omega\right)}
{\Gamma^2+q^2B_{\rm phys}^2},
\qquad
r
=
\frac{\Gamma^2\Omega+qB_{\rm phys}K}
{\Gamma^2+q^2B_{\rm phys}^2}.
\label{eq:s-r-brownian-direct}
\end{equation}
The parameter \(s\) controls radial relaxation, while \(r\) controls rotational drift. In particular, the magnetic field and bath rotation enter these two sectors through different combinations.

We proceed by using
\begin{equation}
\left(
I-\frac{qB_{\rm phys}}{\Gamma}J
\right)^{\top}
\left(
I-\frac{qB_{\rm phys}}{\Gamma}J
\right)
=
\left(
1+\frac{q^2B_{\rm phys}^2}{\Gamma^2}
\right)I
\end{equation}
so that we obtain
\begin{equation}
\mathcal S_{\rm OM}[\bm x]
=
-\frac{1}{4D}
\left(
1+\frac{q^2B_{\rm phys}^2}{\Gamma^2}
\right)
\int_{t_i}^{t_f}dt\,
\left|
\left(
\partial_t+sI+rJ
\right)\bm x
\right|^2.
\label{eq:Seff-reduced-brownian}
\end{equation}
To evaluate the transition probability, we introduce
\begin{equation}
\bm x(t)=U(t,t_i)\bm z(t),
\label{eq:x-U-z}
\end{equation}
where
\begin{equation}
\partial_tU(t,t_i)
=
-(sI+rJ)U(t,t_i),
\qquad
U(t_i,t_i)=I.
\label{eq:U-evolution-brownian}
\end{equation}
Since \(I\) and \(J\) commute,
\begin{equation}
U(t,t_i)
=
e^{-sI(t-t_i)}
e^{-rJ(t-t_i)}.
\label{eq:U-solution-brownian}
\end{equation}
Using Eq.~\eqref{eq:U-evolution-brownian}, one finds
\begin{align}
\left(
\partial_t+sI+rJ
\right)\bm x
&=
\dot U\,\bm z
+U\dot{\bm z}
+(sI+rJ)U\bm z
\nonumber\\
&=
U\dot{\bm z}.
\label{eq:drift-simplification-brownian}
\end{align}
Moreover,
\begin{equation}
U(t,t_i)^{\top}U(t,t_i)
=
e^{-2s(t-t_i)}I,
\end{equation}
because \(e^{-rJ(t-t_i)}\) is orthogonal. Therefore,
\begin{equation}
\mathcal S_{\rm OM}[\bm z]
=
-\frac{1}{4D}
\left(
1+\frac{q^2B_{\rm phys}^2}{\Gamma^2}
\right)
\int_{t_i}^{t_f}dt\,
e^{-2s(t-t_i)}
\left|\dot{\bm z}\right|^2.
\label{eq:SOM-z-final-brownian}
\end{equation}
The boundary conditions are
\begin{equation}
\bm z(t_i)=\bm x_i,
\qquad
\bm z(t_f)
=
U(t_f,t_i)^{-1}\bm x_f
\equiv
\bm z_f.
\label{eq:z-boundary-brownian}
\end{equation}

The Euler--Lagrange equation following from Eq.~\eqref{eq:SOM-z-final-brownian} is
\begin{equation}
\frac{d}{dt}
\left[
e^{-2s(t-t_i)}\dot{\bm z}
\right]
=
0.
\label{eq:EL-brownian}
\end{equation}
For \(s\neq0\), its solution satisfying the initial boundary condition is
\begin{equation}
\bm z(t)
=
\bm x_i
+
\bm C\,
\frac{e^{2s(t-t_i)}-1}{2s},
\label{eq:z-general-brownian}
\end{equation}
where \(\bm C\) is constant. Imposing the final boundary condition gives
\begin{equation}
\bm C
=
\frac{2s}{e^{2s\tau}-1}
\left(
\bm z_f-\bm x_i
\right),
\label{eq:C-solution-brownian}
\end{equation}
and hence
\begin{equation}
\bm z_{\rm cl}(t)
=
\bm x_i
+
\frac{e^{2s(t-t_i)}-1}
{e^{2s\tau}-1}
\left(
\bm z_f-\bm x_i
\right).
\label{eq:z-classical-brownian}
\end{equation}
Transforming back to the original variables, we obtain
\begin{equation}
\bm x_{\rm cl}(t)
=
U(t,t_i)
\left[
\bm x_i
+
\frac{e^{2s(t-t_i)}-1}
{e^{2s\tau}-1}
\left(
U(t_f,t_i)^{-1}\bm x_f-\bm x_i
\right)
\right].
\label{eq:x-classical-brownian}
\end{equation}
The expressions for \(s=0\) are obtained by taking the continuous \(s\to0\) limit.

Evaluating the action on this path gives
\begin{align}
\mathcal S_{\rm cl}
&=
-\frac{1}{4D}
\left(
1+\frac{q^2B_{\rm phys}^2}{\Gamma^2}
\right)
\frac{4s^2}
{\left(e^{2s\tau}-1\right)^2}
\left|
\bm z_f-\bm x_i
\right|^2
\int_{t_i}^{t_f}dt\,
e^{2s(t-t_i)}
\nonumber\\
&=
-\frac{s}
{2D\left(e^{2s\tau}-1\right)}
\left(
1+\frac{q^2B_{\rm phys}^2}{\Gamma^2}
\right)
\left|
\bm z_f-\bm x_i
\right|^2.
\label{eq:Scl-intermediate-brownian}
\end{align}
Since
\begin{equation}
\bm z_f-\bm x_i
=
U(t_f,t_i)^{-1}
\left[
\bm x_f-U(t_f,t_i)\bm x_i
\right]
\label{eq:zf-relation-brownian}
\end{equation}
and
\begin{equation}
\left[U(t_f,t_i)^{-1}\right]^{\top}U(t_f,t_i)^{-1}
=
e^{2s\tau}I,
\end{equation}
the classical action becomes
\begin{equation}
\mathcal S_{\rm cl}
=
-\frac{s}
{2D\left(1-e^{-2s\tau}\right)}
\left(
1+\frac{q^2B_{\rm phys}^2}{\Gamma^2}
\right)
\left|
\bm x_f-U(t_f,t_i)\bm x_i
\right|^2.
\label{eq:Scl-final-brownian}
\end{equation}

The final result for the normalized transition probability is therefore
\begin{equation}
P(\bm x_f,t_f\mid\bm x_i,t_i)
=
\frac{1}{2\pi\sigma^2(\tau)}
\exp\!\left[
-\frac{
\left|
\bm x_f-U(t_f,t_i)\bm x_i
\right|^2
}{
2\sigma^2(\tau)
}
\right],
\label{eq:transition-probability-brownian}
\end{equation}
where
\begin{equation}
\sigma^2(\tau)
=
\frac{D}
{
s\left(
1+\dfrac{q^2B_{\rm phys}^2}{\Gamma^2}
\right)
}
\left(
1-e^{-2s\tau}
\right).
\label{eq:variance-brownian-intermediate}
\end{equation}
Using Eq.~\eqref{eq:s-r-brownian-direct}, this becomes
\begin{equation}
\sigma^2(\tau)
=
\frac{D_{\rm B}}
{\Gamma\left(K-qB_{\rm phys}\Omega\right)}
\left(
1-e^{-2s\tau}
\right).
\label{eq:variance-brownian-direct}
\end{equation}
We note that although Eq.~\eqref{eq:variance-brownian-direct} has been written with an explicit factor \(K-qB_{\rm phys}\Omega\) in the denominator, it remains finite at \(s=0\). Its continuous limit is
\begin{equation}
\left.
\sigma^2(\tau)
\right|_{s=0}
=
\frac{2D_{\rm B}}
{\Gamma^2+q^2B_{\rm phys}^2}\,\tau.
\label{eq:variance-s-zero}
\end{equation}

For any finite \(\tau>0\), the variance in Eq.~\eqref{eq:variance-brownian-intermediate} is positive for all real \(s\). A normalizable stationary distribution exists, however, only when the radial relaxation rate is positive,
\begin{equation}
s>0.
\end{equation}
In fact, if \(s=0\), the variance grows linearly with time, while if \(s<0\), it grows exponentially. Neither case admits a normalizable long-time stationary distribution.
Finally, we see that, since \(\Gamma>0\), the condition $s>0$ is equivalent to
\begin{equation}
K-qB_{\rm phys}\Omega>0.
\label{eq:stationarity-condition-transition}
\end{equation}
In the next section, we recover this condition directly from the associated Fokker--Planck equation.

\section{Fokker--Planck analysis}
\label{sec:fokker-planck}

In this section, we investigate the Fokker-Planck equation for the rotating stochastic Landau model. We compute the probability-current flux in the steady state, and explain physically why the steady-state existence condition obtained from the MSR analysis arises in the Fokker-Planck context, and how the circulating probability flux is established.

\subsection{Steady-state existence condition}

Rewriting Eq.~\eqref{eq:normalized-brownian-sde} in the form
\begin{equation}
\dot{\bm x}
=
-\mathsf A\bm x+\bm\zeta(t),
\qquad
\mathsf A=sI+rJ,
\end{equation}
where \(s\) and \(r\) are defined in Eq.~\eqref{eq:s-r-brownian-direct}, the transformed noise satisfies
\begin{equation}
\left\langle \zeta_i(t)\zeta_j(t')\right\rangle
=
2D_{\rm eff}\delta_{ij}\delta(t-t'),
\qquad
D_{\rm eff}
=
\frac{D\Gamma^2}{\Gamma^2+q^2B_{\rm phys}^2} .
\end{equation}
Following \cite{Risken,vanKampen,Gardiner2004}, the corresponding Fokker--Planck equation \cite{Risken,vanKampen,Gardiner2004} for our system can be easily determined to be
\begin{equation}
\partial_t P(\bm x,t)
=
\nabla_{\bm x}\cdot
\left[
\mathsf A\bm x\,P(\bm x,t)
+
D_{\rm eff}\nabla_{\bm x}P(\bm x,t)
\right].
\label{eq:fp-brownian}
\end{equation}
For a Gaussian steady state with covariance matrix \(\Sigma\), the Lyapunov equation is
\begin{equation}
\mathsf A\Sigma+\Sigma\mathsf A^{\top}=2D_{\rm eff}I.
\label{eq:lyapunov-brownian}
\end{equation}
Since \(\mathsf A=sI+rJ\) and the diffusion is isotropic, the solution is isotropic, and thus
\begin{equation}
\Sigma
=
\frac{D_{\rm eff}}{s}I
=
\frac{D\Gamma}{K-qB_{\rm phys}\Omega}\,I.
\label{eq:Sigma-brownian}
\end{equation}
Therefore, for \(s>0\), the steady-state probability density is
\begin{equation}
P_{\rm ss}(\bm x)
=
\frac{K-qB_{\rm phys}\Omega}{2\pi D\Gamma}
\exp\left[
-\frac{K-qB_{\rm phys}\Omega}{2D\Gamma}
\bm x^{\top}\bm x
\right].
\label{eq:Pss-brownian}
\end{equation}

We can now see that the existence condition in \eqref{eq:stationarity-condition-transition} has a simple physical interpretation. In the absence of the harmonic potential and magnetic field, the overdamped equation (without considering the noise) gives
\begin{equation}
\dot{\bm x}=-\Omega J\bm x.
\end{equation}
For a particle displaced to the right of the origin, \(\bm x=(x,0)\) with \(x>0\), one has
\begin{equation}
J\bm x=(0,-x),
\qquad
\dot{\bm x}=(0,\Omega x).
\end{equation}
Thus, the rotating bath induces an upward tangential drift at this point. If \(q>0\) and \(B_{\rm phys}>0\), the Lorentz force associated with this tangential motion points radially outward,
\begin{equation}
q\bm v\times\bm B
\propto
\hat{\bm y}\times\hat{\bm z}
=
\hat{\bm x}.
\end{equation}
This outward contribution competes with the harmonic potential, whose force points inward, \(-K\bm x\). A steady state exists only when the trap wins this competition, namely when
\begin{equation}
K>qB_{\rm phys}\Omega.
\end{equation}
For the opposite sign of \(qB_{\rm phys}\Omega\), the magnetic deflection of the bath-induced circulation points inward and strengthens the trap. Hence, the product \(qB_{\rm phys}\Omega\) determines whether the magnetic deflection of the rotating-bath drift acts against or with the harmonic confinement.

Interestingly enough, one can have a stochastic steady state even in the absence of an additional trap, as long as the magnetic field, multiplied by the charge, is opposite to the rotation of the medium. In that case, the drift of the particle within the rotating medium leads to a radial Lorentz-force contribution pointing inward and thus confines the particle close to the rotation axis. 

\subsection{Steady-state probability current}

The steady-state probability current \cite{Risken,vanKampen,Gardiner2004} is given by
\begin{equation}
\bm{\mathcal J}_{\rm ss}(\bm x)
=
-\mathsf A\bm x\,P_{\rm ss}(\bm x)
-
D_{\rm eff}\nabla_{\bm x}P_{\rm ss}(\bm x)
=
\left(-\mathsf A+D_{\rm eff}\Sigma^{-1}\right)
\bm x\,P_{\rm ss}(\bm x).
\label{eq:Jss-def-brownian}
\end{equation}
Using \(\Sigma=D_{\rm eff}I/s\) gives
\begin{equation}
D_{\rm eff}\Sigma^{-1}=sI,
\end{equation}
and hence
\begin{equation}
-\mathsf A+D_{\rm eff}\Sigma^{-1}=-rJ.
\end{equation}
Therefore, the steady-state current is purely circulating (and also divergence free):
\begin{equation}
\bm{\mathcal J}_{\rm ss}(\bm x)
=
-rJ\bm x\,P_{\rm ss}(\bm x),
\qquad
r
=
\frac{\Gamma^2\Omega+qB_{\rm phys}K}
{\Gamma^2+q^2B_{\rm phys}^2}.
\label{eq:Jss-final-brownian}
\end{equation}
The result shows that the probability current in the steady state is generically nonzero and moves along contours of constant probability density. 

It is instructive to realize that the two terms in
\begin{equation}
r
=
\frac{\Gamma^2\Omega}
{\Gamma^2+q^2B_{\rm phys}^2}
+
\frac{qB_{\rm phys}K}
{\Gamma^2+q^2B_{\rm phys}^2}
\end{equation}
have distinct physical origins. The first is the circulation induced directly by the rotating bath. If \(B_{\rm phys}=0\), then \(r=\Omega\), so the current co-rotates with the bath. The second is the circulation produced when the magnetic field deflects the inward relaxation generated by the harmonic potential. If \(\Omega=0\), then
\begin{equation}
r
=
\frac{qB_{\rm phys}K}
{\Gamma^2+q^2B_{\rm phys}^2},
\end{equation}
which reduces to the chiral oscillator frequency \(K/(qB_{\rm phys})\) in the conservative massless Landau limit defined when \(\Gamma\to0\). For nonzero \(\Gamma\), the deterministic trajectory spirals while its angular drift is controlled by the frequency given above.

The rotating bath directly induces circulation, while the magnetic field twists the inward relaxation produced by the trap into circulation. If \(qB_{\rm phys}\) and \(\Omega\) have the same sign, these two contributions act in the same direction. The bath-induced circulation and the magnetic deflection of the trap-induced relaxation therefore reinforce each other, enhancing the circulating part of the probability current. If \(qB_{\rm phys}\) and \(\Omega\) have opposite signs, the two effects act against each other. Thus, in this case, the numerator
\begin{equation}
\Gamma^2\Omega+qB_{\rm phys}K
\end{equation}
is reduced, and the rotational drift can even vanish.

We further note that the existence condition for the steady state has the opposite dependence on the relative sign of \(qB_{\rm phys}\) and \(\Omega\). Since the radial relaxation rate is proportional to \(K-qB_{\rm phys}\Omega\), opposite signs of \(qB_{\rm phys}\) and \(\Omega\) imply
\begin{equation}
K-qB_{\rm phys}\Omega
=
K+|qB_{\rm phys}\Omega|>0
\end{equation}
for \(K>0\). Thus, a steady state is guaranteed to exist in this case. Physically, the magnetic deflection of the bath-induced circulation points radially inward, so it strengthens the restoring effect of the harmonic trap rather than opposing it.

To see concretely how probability current circulates in the steady state, place the particle at \(\bm x=(x,0)\) with \(x>0\). The harmonic trap pushes the particle toward the origin, in the \(-\hat{\bm x}\) direction. In the presence of a magnetic field, this inward relaxation is deflected sideways, producing a tangential component of the drift. This is the contribution proportional to \(qB_{\rm phys}K\). Independently, the rotating bath produces a tangential drift. The observed circulation is, however, not simply the rotation frequency of the bath plus the frequency at which the Brownian particle spirals in the presence of the magnetic field.
To see why, set \(K=0\). The deterministic equation becomes
\begin{equation}
\Gamma\dot{\bm x}
-qB_{\rm phys}J\dot{\bm x}
+\Gamma\Omega J\bm x
=
0.
\end{equation}
In the absence of a magnetic field, this gives
\begin{equation}
\dot{\bm x}
=
-\Omega J\bm x
=
(0,\Omega x),
\end{equation}
so the rotating bath produces a purely tangential velocity and the angular drift is simply \(\Omega\).

When \(B_{\rm phys}\neq0\), the magnetic term changes the velocity response. Writing
\begin{equation}
\dot{\bm x}=(v_x,v_y),
\end{equation}
the equation at \(\bm x=(x,0)\) becomes
\begin{equation}
\Gamma(v_x,v_y)
-qB_{\rm phys}(v_y,-v_x)
+(0,-\Gamma\Omega x)
=
0.
\end{equation}
Equivalently,
\begin{equation}
\Gamma v_x-qB_{\rm phys}v_y=0,
\qquad
\Gamma v_y+qB_{\rm phys}v_x-\Gamma\Omega x=0.
\end{equation}
Thus, a tangential response necessarily induces a radial component,
\begin{equation}
v_x=\frac{qB_{\rm phys}}{\Gamma}v_y.
\end{equation}
Solving gives
\begin{equation}
v_y
=
\frac{\Gamma^2\Omega}
{\Gamma^2+q^2B_{\rm phys}^2}\,x,
\qquad
v_x
=
\frac{\Gamma qB_{\rm phys}\Omega}
{\Gamma^2+q^2B_{\rm phys}^2}\,x.
\end{equation}
Therefore, the bath-induced tangential velocity is reduced from \(\Omega x\) to
\begin{equation}
\frac{\Gamma^2\Omega}
{\Gamma^2+q^2B_{\rm phys}^2}\,x.
\end{equation}
The remaining part of the bath response is converted by the magnetic field into radial motion. For \(qB_{\rm phys}\Omega>0\), this radial component points outward at \(\bm x=(x,0)\) and competes with the harmonic trap. This is why the \(\Omega\)-dependent contribution to the rotational drift is not simply \(\Omega\), but rather
\begin{equation}
\frac{\Gamma^2\Omega}
{\Gamma^2+q^2B_{\rm phys}^2}.
\end{equation}
Thus, the magnetic field converts part of the bath-induced tangential response into radial drift, reducing the angular component of that response. The radial drift points outward or inward according to the sign of \(qB_{\rm phys}\Omega\).

Finally, it is useful to distinguish stationary current circulation from irreversibility.  Let us define the physically conjugate reverse process (denoted by the superscript $R$) by
\begin{equation}
\bm x^R(t)
=
\bm x(t_i+t_f-t),
\qquad
B_{\rm phys}^R=-B_{\rm phys},
\qquad
\Omega^R=-\Omega,
\label{eq:physical-reversal-fp}
\end{equation}
while \(K\), \(\Gamma\), and \(D_{\rm B}\) are unchanged. It follows that
\begin{equation}
s^R=s,
\qquad
r^R=-r,
\qquad
D_{\rm eff}^R=D_{\rm eff}.
\end{equation}
Consequently,
\begin{equation}
P_{\rm ss}^R(\bm x)=P_{\rm ss}(\bm x),
\qquad
\bm{\mathcal J}_{\rm ss}^R(\bm x)
=
-\bm{\mathcal J}_{\rm ss}(\bm x).
\end{equation}
Direct substitution into the Gaussian transition kernel gives
\begin{equation}
P_{\rm ss}(\bm x)\,
P_F(\bm x'\mid\bm x;\tau)
=
P_{\rm ss}(\bm x')\,
P_R(\bm x\mid\bm x';\tau),
\label{eq:generalized-detailed-balance-brownian}
\end{equation}
which is the generalized detailed-balance relation for the physically conjugate forward--reverse pair \cite{Crooks1998,AndyLucas2023}. Ordinary same-process detailed balance instead replaces \(P_R\) by \(P_F\), and in the present model it holds if and only if \(r=0\). A nonzero circulating current therefore does not, by itself, imply positive entropy production for the physically conjugate forward--reverse pair considered here, as expected from physical grounds.

\section{Crooks fluctuation theorem and MSR symmetry}
\label{sec:crooks-general}

Having analyzed the steady state through the Fokker--Planck equation, we now show how a local $\mathbb{Z}_2$ symmetry transformation of the MSR path integral gives rise to a boundary functional that controls the ratio of forward and conjugate-reversed path probabilities, which can be directly connected to the Crooks fluctuation theorem \cite{Crooks1998,Crooks1999,JarzynskiWardIdentities_2011}.

\subsection{Symmetry transformation and the MSR action}
Consider a stochastic process with Gaussian noise, described by an MSR action of the form
\begin{equation}
\mathcal S[x,\bar x]
= i\!\int\!dt\,\bar x(t)^{\top}F[x(t)]
-\frac12\!\iint\!dt\,dt'\,
\bar x(t)^{\top}Q(t{-}t')\,\bar x(t'),
\label{eq:MSR_general}
\end{equation}
where $x(t)$ denotes the stochastic variable, $\bar x(t)$ the response field, $F[x]$ the deterministic residual, and $Q$ a symmetric, positive, invertible possibly nonlocal noise kernel. For the coarse-grained time-reversal operation $\Theta$ defined above, we define $F^R[x]\equiv\Theta F[\Theta x]$, where $\Theta$ acts on $x(t)$ and on all time-reversal-odd backgrounds, with $\Theta^2=1$. Thus, $F^R$ is the deterministic residual of the reversed stochastic dynamics, pulled back to the forward time interval. Now, consider the transformation:
\begin{align}
x(t) \to x^{R}(t) = \Theta x(t),
& &
\bar x(t) \to
\bar x^{R}(t)
&=
\Theta \bar x(t)
+\chi(t),
\label{eq:bar_shift_general}
\end{align}
where we define
\begin{align}
\chi(t)
&\equiv
i \Theta \!\int\!dt'\,Q^{-1}(t{-}t')\,\Delta F(t'),
& &
\Delta F \equiv F^{R}-F .
\end{align}
Under this transformation, the action changes as
\begin{equation}
\begin{split}
\mathcal S
\to \mathcal S^{R}
&= \mathcal S +
i\!\int\!dt\,\bar x^{\top}(t)\,\Delta F(t)
+
i\!\int\!dt\,\chi^{\top}(t)\,F[\Theta x(t)]
\\
&\quad
-\!\iint\!dt\,dt'\,
\bar x^{\top}(t)\,\Theta Q(t{-}t')\,\chi(t')
-
\frac12\!\iint\!dt\,dt'\,
\chi^{\top}(t)\,Q(t{-}t')\,\chi(t').
\end{split}
\label{eq:SRminusS-expand}
\end{equation}
By construction,
\begin{equation}
\Theta \int\!dt'\,Q(t{-}t')\,\chi(t')
=
i\,\Delta F[x(t)],
\end{equation}
so the terms linear in $\bar x$ cancel:
\begin{equation}
i\!\int\!dt\,\bar x^{\top}\Delta F
-
\!\iint\!dt\,dt'\,
\bar x^{\top}(t)\,\Theta Q(t{-}t')\,\chi(t')
=0.
\end{equation}
The remaining variation is therefore
\begin{align}
\mathcal S^{R}-\mathcal S
{}&=
i\!\int\!dt\,\chi^{\top}(t)\,F[\Theta x(t)]
-
\frac12\!\iint\!dt\,dt'\,
\chi^{\top}(t)\,Q(t{-}t')\,\chi(t')
\\{}&=
i\!\int\!dt\,\chi^{\top}(t)\,F[\Theta x(t)]
-
\frac{i}{2}\!\int\!dt\,
\chi^{\top}(t)\,\Theta \Delta F[x(t)]
\\{}&=
\frac{i}{2}\!\int\!dt\,\chi^{\top}(t)
\left(F[\Theta x(t)]+\Theta F[x(t)]\right).
\label{eq:SRminusS-reduced}
\end{align}
Using $\Delta F=F^{R}-F$ and the definition of $\chi(t)$, the variation of the action becomes
\begin{equation}
\mathcal S^{R}[x,\bar x]-\mathcal S[x,\bar x]
= -\frac12\!\iint\!dt\,dt'\,
\Big[
F^{R}(t)^{\top}Q^{-1}(t{-}t')F^{R}(t')
- F(t)^{\top}Q^{-1}(t{-}t')F(t')
\Big].
\label{eq:MSR_symmetry_general}
\end{equation}
Next, we show that this term evaluates to a boundary term in the stochastic Landau model. We use the combined operation $\Theta$ defined in Eq.~\eqref{eq:ST-transformation}. Thus, for $t^R=t_i+t_f-t$, we define
\begin{equation}
(\Theta x)(t)=\mathsf Sx(t^R),
\qquad
(\Theta\bar x)(t)=\mathsf S\bar x(t^R).
\label{eq:theta-fields}
\end{equation}
It follows that
\begin{equation}
\partial_t(\Theta x)(t)=-\mathsf S\dot x(t^R).
\end{equation}
Using
\begin{equation}
\mathsf S^{\top}I\mathsf S=I,
\qquad
\mathsf S^{\top}J\mathsf S=-J,
\qquad
J^{\top}=-J,
\qquad
J^2=-I,
\end{equation}
the induced transformation of the $J$-dependent structures is equivalently described by
\begin{equation}
B_{\rm phys}^R=-B_{\rm phys},
\qquad
\Omega^R=-\Omega,
\qquad
\Gamma^R=\Gamma,
\qquad
K^R=K.
\label{eq:theta-parameter-action}
\end{equation} 
Now, for our MSR action 
\begin{equation}
\mathcal S_{\rm MSR}
=
\int_{t_i}^{t_f}dt\,
\left\{
i\bar{\bm x}^{\top}
\left[
\left(I-\frac{qB_{\rm phys}}{\Gamma}J\right)\dot{\bm x}
+\Omega J\bm x
+\frac{K}{\Gamma}\bm x
\right]
-
D\,\bar{\bm x}^{\top}\bar{\bm x}
\right\}
\label{eq:msr-action-brownian-repeat}
\end{equation}
the deterministic operator is 
\begin{equation}
F[x]
=
\left(I-\frac{qB_{\rm phys}}{\Gamma}J\right)\dot x
+\Omega Jx
+\frac{K}{\Gamma}x,
\label{eq:F_brownian}
\end{equation}
so one finds
\begin{equation}
F^R[x]
=
-\left(I+\frac{qB_{\rm phys}}{\Gamma}J\right)\dot x
-\Omega Jx
+\frac{K}{\Gamma}x .
\label{eq:F-brownian-reversed}
\end{equation}
In the MSR action \eqref{eq:msr-action-brownian-repeat}, the noise covariance and its inverse are, respectively,
\begin{equation}
Q(t{-}t')=2D\,I\delta(t-t'),
\qquad
Q^{-1}(t{-}t')=\frac{1}{2D}\,I\delta(t-t').
\label{eq:Q-and-Qinv-divided}
\end{equation}
Here, \(D=D_{\rm B}/\Gamma^2\) is the rescaled noise strength defined in Eq.~\eqref{eq:rescaled-noise}. With this substitution, the MSR action difference in \eqref{eq:MSR_symmetry_general} reduces to the boundary functional
\begin{equation}
\Delta s_{\rm eff}
=
\frac{K-qB_{\rm phys}\Omega}{2D\Gamma}
\,[x^{\top}x]_{t_i}^{t_f}.
\label{eq:deltaS_boundary_brownian}
\end{equation}
Hence, the transformation in \eqref{eq:bar_shift_general} is a generalized-reversal symmetry transformation in the MSR path integral: it maps the forward action to the conjugate-reversed action plus the endpoint functional \eqref{eq:deltaS_boundary_brownian}. In the next section, we interpret this boundary functional through the ratio of forward and conjugate-reversed path probabilities.

\subsection{Connection to Crooks fluctuation theorem}

Consider the transition probability $P_F(b\mid a)$ for stochastic trajectories evolving from an initial configuration $x(t_i)=a$ to a final configuration $x(t_f)=b$. Crooks fluctuation theorem \cite{Crooks1998,Crooks1999} relates the path probability $\mathbb P_F[\gamma\mid a]$ of a forward trajectory $\gamma$ from $a$ to $b$ to the path probability $\mathbb P_R[\gamma^R\mid b^R]$ of its conjugate-reversed trajectory according to
\begin{equation}
\frac{\mathbb P_R[\gamma^R\mid b^R]}
{\mathbb P_F[\gamma\mid a]}
= \exp[-\Delta s_{\rm C}[\gamma]].
\label{eq:crooks_theorem}
\end{equation}
Here, $\Delta s_{\rm C}[\gamma]$ is the entropy produced by this process.

The forward transition probability may be written in its MSR path-integral form as
\begin{equation}
P_F(b\mid a)
= \mathcal N_F
\int_{x(t_i)=a}^{x(t_f)=b}\!\mathcal{D}x\,\mathcal{D}\bar x\;
\Delta_F[x]\,
e^{\mathcal S_F[x,\bar x]},
\qquad
\Delta_F[x]
\equiv
\det\!\left(\frac{\delta F[x]_i}{\delta x_j}\right).
\label{eq:crooks_PI}
\end{equation}
Similarly, the probability of the conjugate-reversed process can be expressed as
\begin{equation}
P_R(a^R\mid b^R)
= \mathcal N_R
\int_{x^R(t_i)=b^R}^{x^R(t_f)=a^R}\!
\mathcal{D}x^R\,\mathcal{D}\bar x^R\;
\Delta_R[x^R]\,
e^{\mathcal S_R[x^R,\bar x^R]}.
\label{eq:reverse_PI}
\end{equation}
One can show that for the additive linear process considered here the following relation holds:
\begin{equation}
\mathcal N_F\,\mathcal{D}x\,\mathcal{D}\bar x\,\Delta_F[x]
=
\mathcal N_R\,\mathcal{D}x^R\,\mathcal{D}\bar x^R\,\Delta_R[x^R].
\label{measureinvariant}
\end{equation}
Thus, using the result from the previous section,
\begin{equation}
\mathcal S_R[x^R,\bar x^R]-\mathcal S_F[x,\bar x]
=\Delta s_{\rm eff}[a,b],
\end{equation}
where $\Delta s_{\rm eff}[a,b]$ depends only on the endpoints, we obtain
\begin{equation}
\frac{P_R(a^R\mid b^R)}{P_F(b\mid a)}
=e^{\Delta s_{\rm eff}[a,b]}.
\label{eq:P_ratio}
\end{equation}
Comparison with \eqref{eq:crooks_theorem} identifies $-\Delta s_{\rm eff}[a,b]$ with $\Delta s_{\rm C}[\gamma]$. This result gives the transformation \eqref{eq:bar_shift_general} a concrete interpretation in the stochastic Landau model: the response-field shift relates the forward and conjugate-reversed MSR weights, with the resulting boundary functional equal to the negative of the entropy entering the Crooks relation.
In the next section, we couple the MSR theory to external sources and show how the local symmetry transformation identified above constrains the source-coupled generating functional and results in MSR Ward identities.


\section{MSR Ward Identities via coupling to sources}
\label{sec:src-integrateout}

\subsection{Deriving the effective action} First, we couple the MSR action to sources such that  the entropy-producing symmetry identified in Sec.~\ref{sec:crooks-general} remains a symmetry. Starting with a general Gaussian MSR action,
\begin{equation}
\mathcal S[x,\bar x]
=
i\!\int\!dt\,\bar x(t)^{\top}(Fx)(t)
-\frac12\!\iint\!dt\,dt'\,
\bar x(t)^{\top}Q(t-t')\bar x(t'),
\label{eq:MSR_Gaussian}
\end{equation}
where $F$ is the linear deterministic operator and $Q$ is the noise kernel. We couple sources by
\begin{equation}
\mathcal S_{\rm src}[\lambda,\bar\lambda;x,\bar x]
=
i\!\int\!dt\,
\Big[
\lambda(t)^{\top}\bar x(t)
+
\bar\lambda(t)^{\top}x(t)
\Big],
\label{eq:Isrc_general}
\end{equation}
and define
\begin{equation}
Z[\lambda,\bar\lambda]
=
\int\!\mathcal D x\,\mathcal D\bar x\;
\exp\!\Big(\mathcal S[x,\bar x]+\mathcal S_{\rm src}[\lambda,\bar\lambda;x,\bar x]\Big)
\equiv
\exp\!\big(\mathcal S_{\rm eff}[\lambda,\bar\lambda]\big).
\label{eq:Z-def}
\end{equation}

The entropy-producing symmetry acts on the MSR fields by time reversal together with a response-field shift. This shift is controlled by the terms that are odd under the coarse-grained time-reversal operation $\Theta$. We have, 
\begin{align}
x(t)
&\mapsto
\Theta x(t),
&
\bar x(t)
&\mapsto
\Theta \bar x(t)
-
\frac{i\Gamma}{D_{\rm B}}
\Theta\Big(I\partial_t+\Omega J\Big)x(t).
\label{eq:xbarx-transform-brownian}
\end{align}
Since $\lambda$ sources $\bar x$ and $\bar\lambda$ sources $x$, invariance of the effective action requires that the sources transform in the following way:
\begin{align}
\lambda(t)
&\mapsto
\Theta \lambda(t),
&
\bar\lambda(t)
&\mapsto
\Theta \bar\lambda(t)
-
\frac{i\Gamma}{D_{\rm B}} \Theta
\Big(I\partial_t+\Omega J\Big)\lambda(t).
\label{eq:lambdabar-transform-brownian}
\end{align}
Applying \eqref{eq:xbarx-transform-brownian} and \eqref{eq:lambdabar-transform-brownian} to \eqref{eq:Isrc_general}, the terms proportional to $\Omega J$ cancel by antisymmetry of $J$. The variation of the source action is therefore
\begin{equation}
\Delta \mathcal S_{\rm src}
=
\frac{\Gamma}{D_{\rm B}}\int dt\,
\Big[
\lambda^{\top}\dot x
+
\dot\lambda^{\top}x
\Big]
=
\frac{\Gamma}{D_{\rm B}}
\int dt\,\partial_t(\lambda^{\top}x).
\label{eq:Isrc-variation-brownian}
\end{equation}
Thus,
\begin{equation}
\Delta \mathcal S_{\rm src}
=
\frac{\Gamma}{D_{\rm B}}
\Big[
\lambda(t)^{\top}x(t)
\Big]_{t_i}^{t_f}.
\label{eq:Isrc-boundary-brownian}
\end{equation}
The source-coupled theory is therefore invariant up to a boundary term. Integrating out $x$ and $\bar x$ gives the quadratic source-only action
\begin{equation}
\begin{split}
\mathcal S_{\rm eff}[\lambda,\bar\lambda]
&=
\iint\!dt\,dt'\,
\bar\lambda(t)^{\top}
K_{\bar\lambda\bar\lambda}(t,t')
\bar\lambda(t')
\\
&\quad
+\iint\!dt\,dt'\,
\lambda(t)^{\top}
K_{\lambda\bar\lambda}(t,t')
\bar\lambda(t')
+
\iint\!dt\,dt'\,
\bar\lambda(t)^{\top}
K_{\bar\lambda\lambda}(t,t')
\lambda(t'),
\end{split}
\label{eq:Ieff-kernels-time}
\end{equation}
with,
\begin{equation}
K_{\bar\lambda\bar\lambda}
=
-\frac12\,G_R*Q*G_A,
\qquad
K_{\bar\lambda\lambda}
=
-\frac{i}{2}G_R,
\qquad
K_{\lambda\bar\lambda}
=
-\frac{i}{2}G_A.
\label{eq:Kernels-general}
\end{equation}
Here,
\begin{equation}
G_R\equiv F^{-1},
\qquad
G_A\equiv (F^\dagger)^{-1}.
\label{eq:GRGA-def-general}
\end{equation}
We use the Fourier convention $f(t)=\int_\omega e^{-i\omega t}f(\omega)$. In frequency space, we have
\begin{equation}
\mathcal S_{\rm eff}[\lambda,\bar\lambda]
=
\int_\omega
\Big[
\bar\lambda(-\omega)^{\top}K_{\bar\lambda\bar\lambda}(\omega)\bar\lambda(\omega)
+\bar\lambda(-\omega)^{\top}K_{\bar\lambda\lambda}(\omega)\lambda(\omega)
+\lambda(-\omega)^{\top}K_{\lambda\bar\lambda}(\omega)\bar\lambda(\omega)
\Big].
\label{eq:Ieff-kernels-omega}
\end{equation}
For the stochastic Landau model with white, isotropic noise, we have
\begin{equation}
Q(t-t')=2D_{\rm B}\,I\,\delta(t-t'),
\label{eq:local-noise-D}
\end{equation}
The kernels become
\begin{equation}
K_{\bar\lambda\bar\lambda}(\omega)
=
-D_{\rm B}\,G_R(\omega)G_A(\omega),
\qquad
K_{\bar\lambda\lambda}(\omega)
=
-\frac{i}{2}G_R(\omega),
\qquad
K_{\lambda\bar\lambda}(\omega)
=
-\frac{i}{2}G_A(\omega).
\label{eq:K-GRGA-brownian}
\end{equation}
And $G_R(\omega)$ for the stochastic Landau model is
\begin{equation}
G_R(\omega)
=
\left[
-i\omega
\left(\Gamma 
I-qB_{\rm phys}J
\right)
+\Gamma \Omega J
+K I
\right]^{-1},
\qquad
G_A(\omega)=G_R(-\omega)^{\top}.
\label{eq:GR-brownian-MSR}
\end{equation}
Note, in writing the symmetry transformations in \eqref{eq:lambdabar-transform-brownian} and $G_R(\omega)$ in \eqref{eq:GR-brownian-MSR}, we have assumed the MSR action to be written with \eqref{eq:massless-stochastic-landau} as the SDE with noise as specified in the equation.

\subsection{Effective action constraints}
\label{sec:FDT-sources}
We now impose the source-symmetry on the source-only effective action. The detailed expansion of the transformed source-only action is given in Appendix~\ref{app:source-variation}. The result is that invariance of the generating functional is equivalent to the vanishing of two kernels,
\begin{equation}
\mathcal K(\omega)=0,
\qquad
\mathcal M(\omega)=0.
\label{eq:Ward-FDT-brownian}
\end{equation}
For the stochastic Landau model, these kernels are
\begin{equation}
\begin{split}
\mathcal K(\omega)
&=
i\Big(G_R(\omega)-G_A(\omega)\Big)
+
2\Gamma \omega\,G_A(\omega)G_R(\omega)
+
2i\Gamma\Omega\,J\,G_A(\omega)G_R(\omega),
\end{split}
\label{eq:Keff-def-brownian}
\end{equation}
and
\begin{equation}
\begin{split}
\mathcal M(\omega)
&=
\frac{i\Gamma \omega}{2}
\Big(G_R(\omega)-G_A(\omega)\Big)
-\frac{\Gamma \Omega}{2}
\Big(G_R(\omega)J-JG_A(\omega)\Big)
\\
&\quad
+
\Gamma^2 G_A(\omega)G_R(\omega)
\Big[
(\omega^2+\Omega^2)I
+
2i\omega\Omega J
\Big]\\
&=\frac{\Gamma}{2}(\omega I + i\Omega J)\mathcal K(\omega)
\end{split}
\label{eq:M-omega-def-brownian}
\end{equation}
Thus, $\mathcal{K}(\omega)=0$ is the response-function identity enforced by the source symmetry of the stochastic Landau model. In the next section, we show that the identity derived here may be interpreted as the high-temperature limit of a fluctuation-dissipation relation derived from a Gibbs state of the form $\rho = Z^{-1}\exp(- \beta(H - \Omega L_z))$, where $L_z$ is the angular momentum operator \cite{salvio2026thermalgaugetheoryrotating, Becattini, Becattini:2020qol, Vilenkin, Haehl:2014zda, GloriosoLiu, CrossleyGloriosoLiu}. This provides a useful interpretation of the Crooks-derived MSR symmetry. In the stochastic Landau model, the symmetry that generates the entropy boundary term is the coarse-grained descendant of a modified KMS condition associated with a rotating thermal state.


\section{Derivation from a rotating KMS condition}
\label{app:twisted-kms}
Thermal states with chemical potential and rotation have been much discussed in the literature \cite{Haag:1967sg,Becattini,Becattini:2020qol,Vilenkin,AMBRUS2014296,88mq-j66r}.  Here, we derive the non-relativistic fluctuation-dissipation relations implied by the `twisted' KMS condition obeyed by a thermal state with conserved (non-vanishing) angular momentum \cite{salvio2026thermalgaugetheoryrotating}. Upon identifying the strength of the noise with an effective temperature scale, the resulting relation matches precisely the symmetry constraint derived at the level of the MSR effective action. Thus, we may conclude that the MSR Ward identity derived via Crooks theorem is consistent with an underlying microscopic symmetry encoded in the twisted KMS condition \cite{GloriosoLiu, Haehl:2014zda, Haehl:2015pja}.

For completeness, we now derive the fluctuation-dissipation-type relation associated
with the rotated thermal density matrix:
\begin{equation}
\rho_{\beta,\theta}
=
\frac{1}{Z_{\beta,\theta}}
e^{-\beta (H-\theta L_z)},
\qquad
Z_{\beta,\theta}
=
\Tr \!\left(e^{-\beta (H-\theta L_z)}\right).
\end{equation}
We assume that $H$ is time independent and that $L_z$ generates an $SO(2)$
symmetry, so that $[H,L_z]=0$. The Heisenberg operators are $x_i(t)=e^{iHt}x_i e^{-iHt}$ and the action of $L_z$ on $x_i$ is
\begin{equation}
[L_z,x_i]=iJ_{i\ell}x_\ell,
\qquad
J^2=-I.
\end{equation}
The thermal two-point function is
\begin{equation}
W_{ij}(t,t')
=
\langle x_i(t)x_j(t')\rangle_{\beta,\theta}
=
\Tr \!\left(
\rho_{\beta,\theta}\,x_i(t)x_j(t')
\right).
\end{equation}
By cyclicity of the trace,
\begin{equation}
\begin{split}
W_{ij}(t,t')=
\frac{1}{Z_{\beta,\theta}}
\Tr \!\left(
x_j(t')e^{-\beta H}e^{\beta\theta L_z}x_i(t)
\right),
\end{split}
\label{eq:app-cyclic-KMS-start}
\end{equation}
where we used $[H,L_z]=0$ to factorize the exponential. We now insert
$e^{\beta\theta L_z}e^{-\beta\theta L_z}=1$ and
$e^{\beta H}e^{-\beta H}=1$ so that the thermal factors conjugate
$x_i(t)$. This gives
\begin{equation}
W_{ij}(t,t') =
\Tr\!\left[
\rho_{\beta,\theta}\,
x_j(t')
e^{\beta\theta L_z}
e^{-\beta H}x_i(t)e^{\beta H}
e^{-\beta\theta L_z}
\right].
\label{eq:app-trace-manipulation}
\end{equation}
The Hamiltonian conjugation shifts the Heisenberg operator in imaginary time $e^{-\beta H}x_i(t)e^{\beta H}=x_i(t+i\beta)$ and the $SO(2)$ conjugation gives
\begin{equation}
e^{\beta\theta L_z}x_i(t+i\beta)e^{-\beta\theta L_z}
=
\left(e^{i\beta\theta J}\right)_{i\ell}
x_\ell(t+i\beta),
\label{eq:app-angular-momentum-conjugation}
\end{equation}
which follows directly from $[L_z,x_i]=iJ_{i\ell}x_\ell$. Therefore,
\begin{equation}
W_{ij}(t,t')
=
\left(e^{i\beta\theta J}\right)_{i\ell}
W_{j\ell}(t',t+i\beta).
\label{eq:app-twisted-KMS-time}
\end{equation}
This is the twisted KMS condition in the time domain. Assuming time-translation invariance and taking Fourier transforms, one finds
\begin{equation}
W(\omega)
=
e^{\beta\omega}
e^{i\beta\theta J}
W^{\top}(-\omega),
\qquad
W^{\top}(-\omega)
=
e^{-\beta\omega}
e^{-i\beta\theta J}
W(\omega).
\label{eq:app-twisted-KMS-frequency}
\end{equation}
We now define the antisymmetrized and symmetrized correlators by
\begin{equation}
G(\omega)
=
\frac{1}{2}
\left[
W(\omega)-W^{\top}(-\omega)
\right],
\qquad
G_S(\omega)
=
\frac{1}{2}
\left[
W(\omega)+W^{\top}(-\omega)
\right].
\label{eq:app-G-GS-def}
\end{equation}
Introducing $\mathcal Q_{\beta,\theta}(\omega):=e^{-\beta\omega}e^{-i\beta\theta J}$, we rewrite
\begin{equation}
G(\omega)
=
\frac{1}{2}\left[I-\mathcal Q_{\beta,\theta}(\omega)\right]W(\omega),
\qquad
G_S(\omega)
=
\frac{1}{2}\left[I+\mathcal Q_{\beta,\theta}(\omega)\right]W(\omega),
\end{equation}
and therefore
\begin{equation}
G_S(\omega)
=
\left[I+\mathcal Q_{\beta,\theta}(\omega)\right]
\left[I-\mathcal Q_{\beta,\theta}(\omega)\right]^{-1}
G(\omega),
\qquad
\det\!\left[I-\mathcal Q_{\beta,\theta}(\omega)\right]\neq0.
\label{eq:app-FDT-Q-form}
\end{equation}
It remains to simplify the matrix factor. Using $J^2=-I$, we write
\begin{equation}
\mathcal Q_{\beta,\theta}(\omega)
=
u_\omega\left(c_\theta I-is_\theta J\right),
\qquad
u_\omega=e^{-\beta\omega},\quad
c_\theta=\cosh(\beta\theta),\quad
s_\theta=\sinh(\beta\theta).
\end{equation}
Then
\begin{equation}
\left(I+\mathcal Q_{\beta,\theta}(\omega)\right)
\left(I-\mathcal Q_{\beta,\theta}(\omega)\right)^{-1}
=
\frac{(1-u_\omega^2)I-2iu_\omega s_\theta J}
{1-2u_\omega c_\theta+u_\omega^2}.
\label{eq:app-Q-product-r-c-s}
\end{equation}
Multiplying numerator and denominator by $e^{\beta\omega}$ converts this to
hyperbolic functions of $\beta\omega$:
\begin{equation}
\left(I+\mathcal Q_{\beta,\theta}(\omega)\right)
\left(I-\mathcal Q_{\beta,\theta}(\omega)\right)^{-1}
=
\frac{
\sinh(\beta\omega)I
-
i\sinh(\beta\theta) J
}{
\cosh(\beta\omega)-\cosh(\beta\theta)
}.
\label{eq:app-Q-product-final}
\end{equation}
Thus, the exact twisted fluctuation-dissipation-type relation is
\begin{equation}
G_S(\omega)
=
\frac{
\sinh(\beta\omega)I
-
i\sinh(\beta\theta) J
}{
\cosh(\beta\omega)-\cosh(\beta\theta)
}
G(\omega),
\qquad
\omega^2\neq\theta^2.
\label{eq:app-twisted-FDT-exact}
\end{equation}
One may check that this reduces to the ordinary thermal
fluctuation-dissipation relation for $\theta=0$.

Finally, we take the high-temperature limit of the full twisted relation.
For small $\beta$,
\begin{equation}
\sinh(\beta\omega)=\beta\omega+O(\beta^3),
\qquad
\sinh(\beta\theta)=\beta\theta+O(\beta^3),
\end{equation}
and
\begin{equation}
\cosh(\beta\omega)-\cosh(\beta\theta)
=
\frac{\beta^2}{2}
\left(\omega^2-\theta^2\right)
+
O(\beta^4).
\end{equation}
Eq.~\eqref{eq:app-twisted-FDT-exact} becomes
\begin{equation}
G_S(\omega)
=
\left[
2T
\frac{
\omega I-i\theta J
}{
\omega^2-\theta^2
}
+
O(T^{-1})
\right]
G(\omega),
\qquad
\omega^2\neq\theta^2.
\label{eq:app-twisted-FDT-high-T}
\end{equation}
Equivalently,
\begin{equation}
G(\omega)
=
\left[
\frac{1}{2T}
(\omega I+i\theta J)
+
O(T^{-3})
\right]
G_S(\omega).
\label{eq:app-twisted-FDT-high-T-solved}
\end{equation}
This is the high temperature limit of the fluctuation-dissipation relation implied by the modified KMS condition.

We now relate the antisymmetrized correlator appearing in
Eq.~\eqref{eq:app-twisted-FDT-high-T-solved} to the usual retarded and
advanced quantum Green functions.
We define,
\begin{equation}
G^R_{ij}(t)
=
i\Theta_{\rm H}(t)
\left\langle
[x_i(t),x_j(0)]
\right\rangle_{\beta,\theta},
\qquad
G^A_{ij}(t)
=
-i\Theta_{\rm H}(-t)
\left\langle
[x_i(t),x_j(0)]
\right\rangle_{\beta,\theta}.
\end{equation}
Thus, \eqref{eq:app-twisted-FDT-high-T-solved} can be rewritten as
\begin{equation}
G^R(\omega)-G^A(\omega)
=
\left[
\frac{i}{T}
(\omega I+i\theta J)
+
O(T^{-3})
\right]
G_S(\omega).
\label{eq:app-quantum-retarded-FDT}
\end{equation}
Now, consider the kernel from \eqref{eq:Keff-def-brownian}. Using $x(\omega)=G_R(\omega)\xi(\omega)$ for the stochastic dynamics with $\langle \xi(\omega)\xi^{\top}(\omega')\rangle = 2D_{\rm B}(2\pi)\delta(\omega+\omega')\,I$ and defining the symmetrized correlator as in \eqref{eq:app-G-GS-def}, we can rewrite $\mathcal{K}(\omega)$ as:
\begin{equation}
G_R(\omega)-G_A(\omega)=\frac{i\Gamma}{D_{\rm B}}(\omega I+i \Omega J) \,G_S(\omega)
\label{eq:FDT-MSR}
\end{equation}
Comparing \eqref{eq:app-quantum-retarded-FDT} and \eqref{eq:FDT-MSR}, we see that the two expressions are equivalent upon identifying $\Omega$ as the ``charge'' of the angular momentum operator $\theta$ in the Gibbs state and identifying $D_{\rm B}= \Gamma \,T$, which is the standard Einstein relation in our units. Thus, we see that the structure of the response-function relations implied by the MSR Ward identity is identical to the high-temperature limit of the fluctuation-dissipation theorem obtained from a rotating thermal Gibbs state. Upon matching coefficients, we see that noise strength $D_{\rm B}$ plays the role of a scaled-temperature in the stochastic theory. Therefore, the Crooks-time reversal implementing symmetry transformation of \eqref{eq:bar_shift_general} may be interpreted as the coarse-grained descendant of a modified dynamical KMS symmetry. We interpret this as saying that the stochastic Landau model describes a Brownian particle in rotating thermal equilibrium.


\section{Conclusions and outlook}
\label{sec:conclusions}

In this work, we studied a rotating stochastic analog of the planar Landau model as an exactly solvable setting that allows one to determine how much of the structure of fluctuation--response relations emerges from implementing time-reversal symmetry in the coarse-grained stochastic dynamics and which require additional input. The model describes an overdamped charged Brownian particle coupled dissipatively to a rotating environment and subject to a magnetic field, whose Gaussian steady state supports a generically nonzero circulating probability current. Thus, this system provides a simple example in which time reversal, entropy production, and response constraints can be studied in the presence of steady-state circulating currents.

Motivated by previous work \cite{Mullins:2025vqa}, we identified an emergent local \(\mathbb{Z}_2\) symmetry transformation that implements the physically reversed stochastic dynamics, which here includes reversal of the time-reversal-odd backgrounds \(B_{\rm phys}\) and \(\Omega\). Under this transformation, the MSR action changes only by a boundary term. In fact, for the stochastic Landau model, this boundary contribution is precisely the endpoint entropy associated with the steady-state distribution. Thus, the Crooks entropy is realized at the level of the stochastic path integral through a transformation of the response field. 

After coupling the MSR theory to external sources, the same Crooks-compatible transformation constrains the source-dependent generating functional. In the linear Gaussian theory considered here, these constraints take the form of response-function Ward identities derived entirely within the coarse-grained stochastic description. However, they should not be interpreted by themselves as a complete thermal fluctuation--dissipation theorem because the symmetry does not fix the absolute normalization of the noise kernel. The thermal interpretation therefore requires an additional matching condition.

In fact, if the stochastic model is regarded as the high-temperature effective description of a particle coupled to a rotating thermal environment, the appropriate microscopic ensemble is a rotating Gibbs state of the form
\[
\rho_{\beta,\Omega}
=
\frac{1}{Z}\exp[-\beta(H-\Omega L_z)] .
\]
The corresponding twisted KMS condition gives the thermal fluctuation--dissipation relation. Once the Einstein relation is used to identify the noise strength with the appropriate thermal scale, the MSR Ward identities are found to coincide with the high-temperature limit of the rotating-frame fluctuation--dissipation relations. In this sense, the Crooks fluctuation theorem gives the coarse-grained time-reversal symmetry a path-probability interpretation, the associated MSR transformation yields the Ward identities, and the twisted KMS construction provides their  thermal interpretation.

These results suggest several directions for future work. Because the Crooks fluctuation theorem also applies in nonequilibrium settings, it would be interesting to construct examples in which the resulting entropy-generating symmetry can be interpreted as the descendant of a microscopic symmetry of a specific driven open quantum system or of Lindbladian evolution. A related direction is to investigate what happens when the assumptions underlying the Gaussian Markovian noise used here are relaxed. In the present model, the bath is represented by local Gaussian fluctuations, as is natural when many environmental degrees of freedom contribute independently to the coarse-grained noise. More generally, however, the passage from an open quantum system to a stochastic or hydrodynamic effective theory depends on the structure of the bath, the decoherence mechanism, and the truncation used to define the late-time variables \cite{FeynmanVernon1963,CaldeiraLeggett1983,BreuerPetruccione,ZhouHippertMullinsNoronha,GoriniKossakowskiSudarshan1976,Lindblad1976}. It would therefore be interesting to extend the present construction to baths that generate colored Gaussian noise, multiplicative noise, or genuinely non-Gaussian fluctuations. In such cases, entropy production need not be encoded only in quadratic MSR terms, and the entropy-generating transformation may act nonlinearly on the response fields or on an enlarged set of auxiliary fields \cite{AronBiroliCugliandolo2010,StapmannsBoltz2018}. The resulting Ward identities could then impose higher-point fluctuation--response relations whose microscopic origin may be traceable to symmetries of the underlying open-system or Lindbladian dynamics.

Finally, a natural application of this perspective is to stochastic descriptions of spin polarization in rotating media. In this context, the Barnett effect is the magnetization, or spin polarization, induced by mechanical rotation. In relativistic spin hydrodynamics \cite{Florkowski:2017ruc}, the analogous statement is that vorticity and rotation act as sources for spin polarization, with angular-momentum conservation allowing interconversion between orbital and spin angular momentum \cite{Buzzegoli2023Barnett,Huang2024SpinHydro,Florkowski2024SpinHydro}. In fact, recent formulations of spin magnetohydrodynamics make this connection explicit through relativistic analogs of the Einstein--de Haas and Barnett effects \cite{Bhadury2022SpinMHD,Kiamari2024SpinfulVorticalMHD}. A stochastic extension of such theories would naturally include fluctuations of the spin density around its Barnett-polarized steady value, analogous to the noise terms of fluctuating hydrodynamics and constrained, near equilibrium, by fluctuation--dissipation relations \cite{MuraseHirano2013}. Applying the same Crooks construction to the enlarged MSR action is expected to yield Ward identities that may constrain not only the Barnett susceptibility, namely the spin response to rotation, but also the associated spin fluctuations and mixed spin-orbital response functions. This may provide a controlled route to fluctuation--dissipation relations for Barnett-type response in coarse-grained spin hydrodynamics. We leave this direction for future work.

\section*{Acknowledgements} D.K. and J.N. are partly supported by the U.S. Department of Energy,
Office of Science, Office for Nuclear Physics under Award No. DE-SC0023861. N.M. was partially supported by
U.S. Department of Energy, Office of Nuclear Physics,
Contract DE-FG02-03ER41260. M.H. was supported by the Brazilian National Council for Scientific and Technological Development (CNPq) under process No. 313638/2025-0. M.H. also acknowledges the hospitality of the Illinois Center for Advanced Studies of the Universe, at the University of Illinois Urbana-Champaign, during part of this work.

\appendix

\section{Invariance of the MSR Jacobian under the coarse-grained time-reversal operation}
\label{app:msr-jacobian}

In this appendix, we compute the MSR Jacobian for the rotating stochastic Landau model in the Stratonovich prescription and show that it is invariant under the coarse-grained time-reversal operation used in the main text. We start from the deterministic operator in \eqref{eq:F_brownian}, written as
\begin{equation}
F[\bm x](t)
=
C\,\dot{\bm x}(t)+M\bm x(t),
\label{eq:F-CM-form}
\end{equation}
with
\begin{equation}
C
=
I-\frac{qB_{\rm phys}}{\Gamma}J,
\qquad
M
=
\frac{K}{\Gamma}I
+
\Omega J .
\label{eq:C-M-brownian}
\end{equation}
The MSR Jacobian is
\begin{equation}
\Delta
=
\det\!\left(
\frac{\delta F_{i}(t)}{\delta x_j(t')}
\right)
=
\det\!\left(C\partial_t+M\right).
\label{eq:Delta-def-brownian}
\end{equation}
Discretize time as
\begin{equation}
t_n=t_i+n\varepsilon,
\qquad
n=0,1,\ldots,N,
\qquad
N\varepsilon=\tau .
\end{equation}
In the Stratonovich prescription,
\begin{equation}
F_{n}
=
C\,\frac{\bm x_n-\bm x_{n-1}}{\varepsilon}
+
M.
\label{eq:F-discrete-stratonovich}
\end{equation}
Treating $\bm x_0$ as fixed and differentiating with respect to $\bm x_1,\ldots,\bm x_N$, the Jacobian matrix is lower bidiagonal. Therefore, only the diagonal blocks contribute:
\begin{equation}
\Delta
=
\prod_{n=1}^{N}
\det\!\left(
\frac{C}{\varepsilon}
+\frac12M
\right).
\label{eq:Delta-product-brownian}
\end{equation}
Factoring out $C/\varepsilon$ from each block gives
\begin{equation}
\Delta
=
\prod_{n=1}^{N}
\det\!\left(\frac{C}{\varepsilon}\right)
\det\!\left(
I+\frac{\varepsilon}{2}C^{-1}M
\right).
\label{eq:Delta-factorized-brownian}
\end{equation}
Using
\begin{equation}
\det C
=
1+\frac{q^2B_{\rm phys}^2}{\Gamma^2},
\qquad
C^{-1}M
=
sI+r J ,
\label{eq:CinvM-quote}
\end{equation}
where \(s\) and \(r\) are the Brownian coefficients defined in \eqref{eq:s-r-brownian-direct}, we find
\begin{equation}
\det\!\left(
I+\frac{\varepsilon}{2}C^{-1}M
\right)
=
\left(1+\frac{\varepsilon s}{2}\right)^2
+
\left(
\frac{\varepsilon}{2}r
\right)^2 .
\label{eq:det-small-block-brownian}
\end{equation}
Expanding for small $\varepsilon$,
\begin{equation}
\left(1+\frac{\varepsilon s}{2}\right)^2
+
\left(
\frac{\varepsilon}{2}r
\right)^2
=
1+\varepsilon s+\mathcal O(\varepsilon^2).
\label{eq:block-expansion-brownian}
\end{equation}
Thus,
\begin{equation}
\ln\Delta
=
N\ln\!\left[
\frac{1+\frac{q^2B_{\rm phys}^2}{\Gamma^2}}{\varepsilon^2}
\right]
+
s\tau
+
\mathcal O(\varepsilon),
\end{equation}
or
\begin{equation}
\Delta
=
\mathcal N_\varepsilon e^{s\tau},
\qquad
\mathcal N_\varepsilon
=
\left[
\frac{1+\frac{q^2B_{\rm phys}^2}{\Gamma^2}}{\varepsilon^2}
\right]^N .
\label{eq:Delta-final-brownian}
\end{equation}
The factor $\mathcal N_\varepsilon$ is divergent but independent of the path and is absorbed into the normalization of the MSR measure.

Equation \eqref{eq:Delta-final-brownian} shows that the Jacobian is invariant under the coarse-grained time-reversal operation used in the main text. In fact, under time reversal,
\begin{equation}
B_{\rm phys}\to -B_{\rm phys},
\qquad
\Omega\to -\Omega,
\end{equation}
while $\Gamma$ and $K$ are invariant. However, the finite Jacobian depends on these parameters only through
\begin{equation}
s
=
\frac{\Gamma\bigl(K-qB_{\rm phys}\Omega\bigr)}
{\Gamma^2+q^2B_{\rm phys}^2},
\end{equation}
which is unchanged under this transformation. The divergent normalization $\mathcal N_\varepsilon$ is also unchanged, since it depends only on $B_{\rm phys}^2$. 

\section{Invariance of the effective action}
\label{app:source-variation}
In this appendix, we compute the frequency-space variation of the source-only effective action and derive the form of the kernels $\mathcal{M(\omega)}$ and $\mathcal{K(\omega)}$ stated in Sec.~\ref{sec:FDT-sources}. These kernels must vanish for the effective action to remain invariant under the entropy-generating symmetry.
We begin with,
\begin{equation}
\mathcal S_{\rm eff}[\lambda,\bar\lambda]
=
\mathcal S_{\bar\lambda\bar\lambda}
+
\mathcal S_{\lambda\bar\lambda}
+
\mathcal S_{\bar\lambda\lambda},
\end{equation}
where
\begin{align}
\mathcal S_{\bar\lambda\bar\lambda}
&=
\int_\omega
\bar\lambda(-\omega)^{\top}
K_{\bar\lambda\bar\lambda}(\omega)
\bar\lambda(\omega),
\\
\mathcal S_{\lambda\bar\lambda}
&=
\int_\omega
\lambda(-\omega)^{\top}
K_{\lambda\bar\lambda}(\omega)
\bar\lambda(\omega),
\\
\mathcal S_{\bar\lambda\lambda}
&=
\int_\omega
\bar\lambda(-\omega)^{\top}
K_{\bar\lambda\lambda}(\omega)
\lambda(\omega).
\end{align}
The source transformation is \eqref{eq:lambdabar-transform-brownian}, in frequency space, becomes
\begin{equation}
\lambda(-\omega)
\mapsto
\Theta \lambda(\omega),
\qquad
\bar\lambda(-\omega)
\mapsto
\Theta \bar\lambda(\omega)
-
\frac{\Gamma}{D_{\rm B}}\Theta
\big(\omega I+i\Omega J\big)\lambda(\omega),
\label{eq:app-source-transform}
\end{equation}
We define the time-reversed kernels by $
\widetilde K_{XY}(\omega)
\equiv
K_{XY}(-\omega;-B_{\rm phys},-\Omega)$ where $X,Y\in\{\lambda,\bar\lambda\}$. It is also useful to define $ \Delta K_{XY}(\omega)
\equiv
\widetilde K_{XY}(\omega)-K_{XY}(\omega)$.
Using \eqref{eq:app-source-transform}, the variation of the source-only effective action may be organized by powers of $1/D_{\rm B}$:
\begin{equation}
\Delta \mathcal S_{\rm eff}
=
\Delta \mathcal S_{\rm eff}^{(0)}
+
\Delta \mathcal S_{\rm eff}^{(1)}
+
\Delta \mathcal S_{\rm eff}^{(2)}.
\end{equation}

At zeroth order,
\begin{equation}
\begin{split}
\Delta \mathcal S_{\rm eff}^{(0)}
=
\int_\omega
\bigg[
&\bar\lambda(-\omega)^{\top}
\Delta K_{\bar\lambda\bar\lambda}(\omega)
\bar\lambda(\omega)
+
\lambda(-\omega)^{\top}
\Delta K_{\lambda\bar\lambda}(\omega)
\bar\lambda(\omega)
\\
&\qquad
+
\bar\lambda(-\omega)^{\top}
\Delta K_{\bar\lambda\lambda}(\omega)
\lambda(\omega)
\bigg].
\end{split}
\label{eq:DeltaI0-brownian}
\end{equation}

At first order,
\begin{equation}
\begin{split}
\Delta \mathcal S_{\rm eff}^{(1)}
&=
\int_\omega
\bigg\{
\frac{-\Gamma \omega}{D_{\rm B}}
\bigg[
\bar\lambda(-\omega)^{\top}
\widetilde K_{\bar\lambda\bar\lambda}(\omega)
\lambda(\omega)
-
\lambda(-\omega)^{\top}
\widetilde K_{\bar\lambda\bar\lambda}(\omega)
\bar\lambda(\omega)
\\
&\hspace{8em}
+
\lambda(-\omega)^{\top}
\widetilde K_{\lambda\bar\lambda}(\omega)
\lambda(\omega)
-
\lambda(-\omega)^{\top}
\widetilde K_{\bar\lambda\lambda}(\omega)
\lambda(\omega)
\bigg]
\\
&\quad
-i\frac{\Gamma \Omega}{D_{\rm B}}
\bigg[
\bar\lambda(-\omega)^{\top}
\widetilde K_{\bar\lambda\bar\lambda}(\omega)
J\lambda(\omega)
-
\lambda(-\omega)^{\top}
J\widetilde K_{\bar\lambda\bar\lambda}(\omega)
\bar\lambda(\omega)
\\
&\hspace{8em}
+
\lambda(-\omega)^{\top}
\widetilde K_{\lambda \bar \lambda}(\omega)
J\lambda(\omega)
-
\lambda(-\omega)^{\top}
J\widetilde K_{\bar \lambda \lambda}(\omega)
\lambda(\omega)
\bigg]
\bigg\}.
\end{split}
\label{eq:DeltaI1-brownian}
\end{equation}

At second order,
\begin{equation}
\Delta \mathcal S_{\rm eff}^{(2)}
=
\int_\omega
\frac{\Gamma^2}{D_{\rm B}^2}
\lambda(-\omega)^{\top}
\big[-\omega I-i\Omega J\big]
\widetilde K_{\bar\lambda\bar\lambda}(\omega)
\big[\omega I+i\Omega J\big]
\lambda(\omega).
\label{eq:DeltaI2-brownian}
\end{equation}

For local, isotropic diagonal noise in \eqref{eq:local-noise-D} and the source kernels of \eqref{eq:K-GRGA-brownian} and using the fact that $G_R(-\omega, -B_{\rm phys}, -\Omega)=G_R(-\omega, B_{\rm phys},\Omega)^{\top}=G_A(\omega, B_{\rm phys}, \Omega)$ in the rotating stochastic Landau model, we obtain:
\begin{equation}
\Delta \mathcal S_{\rm eff}
=
\int_\omega
\lambda(-\omega)^{\top}
\mathcal K(\omega)
\bar\lambda(\omega)
+
\int_\omega
\frac{1}{D_{\rm B}}
\lambda(-\omega)^{\top}
\mathcal M(\omega)
\lambda(\omega).
\label{eq:DeltaI-final-brownian}
\end{equation}
The kernels are
\begin{equation}
\begin{split}
\mathcal K(\omega)
&=
i\Big(G_R(\omega)-G_A(\omega)\Big)
+
2\Gamma \omega\,G_A(\omega)G_R(\omega)
+
2i\Gamma\Omega\,J\,G_A(\omega)G_R(\omega),
\end{split}
\label{eq:app-Keff-def-brownian}
\end{equation}
and
\begin{equation}
\begin{split}
\mathcal M(\omega)
&=
\frac{i\Gamma \omega}{2}
\Big(G_R(\omega)-G_A(\omega)\Big)
-\frac{\Gamma \Omega}{2}
\Big(G_R(\omega)J-JG_A(\omega)\Big)
\\
&\quad
+
\Gamma^2 G_A(\omega)G_R(\omega)
\Big[
(\omega^2+\Omega^2)I
+
2i\omega\Omega J
\Big].
\end{split}
\label{eq:app-M-omega-def-brownian}
\end{equation}
Note, in the rotating stochastic Landau model, \(G_R\) and \(G_A\) are both linear combinations of \(I\) and \(J\). All such matrices commute with each other. Moreover, $(\omega I + i\Omega J)^2 = (\omega^2 + \Omega^2)I + 2i\omega\Omega J$. Using this, we may rewrite
\begin{equation}
\mathcal M(\omega)
=
\frac{\Gamma}{2}
(\omega I+i\Omega J)
\mathcal K(\omega).
\end{equation}
Therefore, invariance of the source-only effective action is equivalent to
\begin{equation}
\mathcal K(\omega)=0,
\qquad
\mathcal M(\omega)=0
\label{kernels}
\end{equation}
These are the MSR Ward identities stated in Sec.~\ref{sec:FDT-sources}.

\bibliographystyle{apsrev4-1}
\bibliography{refs}

\end{document}